\documentclass{aa} 
\usepackage{graphicx, amssymb,epsfig} %Modif BS le 10/09/2002 001 : Ca marche sur le Mac !
\usepackage{mathptm}
\usepackage{times}
\usepackage{txfonts}
\usepackage{multirow}
\usepackage{natbib}
\usepackage{longtable}
\usepackage{color}
\usepackage[outercaption]{sidecap}
\usepackage[usenames,dvipsnames]{xcolor}
\newcommand{\blau}{\color{blue}}%
\newcommand\T{\rule{0pt}{2.6ex}}
\newcommand\B{\rule[-1.5ex]{0pt}{0pt}}

\begin{document} 
\title{Physical structure of the photodissociation regions in NGC 7023}
\subtitle{Observations of gas and dust emission with {\it Herschel}\thanks{{\it Herschel} is an ESA space observatory with science instruments provided by European-led Principal Investigator consortia and with important participation from NASA.}}
\author{M. K{\"o}hler\inst{1}\and
              E. Habart\inst{1}\and 
              H. Arab\inst{1}\and 
              J. Bernard-Salas\inst{1,2}\and
              H. Ayasso\inst{1}\and 
              A. Abergel\inst{1}\and
              A. Zavagno\inst{3}\and
              E. Polehampton\inst{4,5}\and
              M.~H.~D. van der Wiel\inst{5}\and 
              D.~A.~Naylor\inst{5}\and 
              G.~Makiwa\inst{5}\and    
              K. Dassas\inst{1}\and           
              C. Joblin\inst{6,7}\and
              P. Pilleri\inst{8}\and
              O. Bern\'e\inst{6,7}\and
              A. Fuente\inst{9}\and
              M. Gerin\inst{10}\and
              J. R. Goicoechea\inst{11}\and
              D. Teyssier\inst{12}}  
              
  \institute{Institut d'Astrophysique Spatiale (IAS), Universit\'e Paris Sud \& CNRS, B\^at. 121, 
       Orsay 91405, France
       \and
       Department of Physical Sciences, The Open University, Milton Keynes MK7 6AA, UK
       \and
       Laboratoire d'Astrophysique de Marseille (UMR 6110 CNRS and Universit\'e de Provence), 38 avenue Joliot Curie, 13388 Marseille CEDEX 13, France
       \and
       RAL Space, Rutherford Appleton Laboratory, Didcot OX11 0QX, UK 
       \and
       Institute for Space Imaging Science, Department of Physics \& Astronomy, University of Lethbridge, Lethbridge, AB T1K3M4, Canada
       \and
      Universit\'e de Toulouse, UPS-OMP, IRAP, 31400 Toulouse, France
       \and
        CNRS, IRAP, 9 Av. Colonel Roche, BP 44346, 31028 Toulouse Cedex 4, France
       \and
       Los Alamos National Laboratory, P.O. Box 1663, Los Alamos, NM 87545, USA
       \and
       Observatorio Astron\'omico Nacional (OAN,IGN), Apdo 112, E-28803 Alcal\'a de Henares, Spain
       \and
       LERMA, Observatoire de Paris, 61 Av. de l'Observatoire, 75014 Paris, France
       \and
       Departamento de Astrof\'{\i}sica. Centro de Astrobiolog\'{\i}a. CSIC-INTA.
       Carretera de Ajalvir, Km 4. Torrej\'on de Ardoz, 28850, Madrid, Spain
       \and
       European Space Astronomy Centre (ESAC), P.O. Box 78, 28691 Villanueva de la Ca\~{n}ada, Madrid, Spain} 
 
\date{Received ..; accepted ..} 

\abstract
  % Context
  {The determination of the physical conditions in molecular clouds is a key step towards our understanding of their formation and evolution of associated star formation.
We investigate the density, temperature, and column density of both dust and gas in the photodissociation regions (PDRs) located at the interface between the atomic and cold molecular gas of the NGC 7023 reflection nebula. 
We study how young stars affect the gas and dust in their environment.}
  % Aims
  {Several \textit{Herschel} Space Telescope programs provide a wealth of spatial and spectral information of dust and gas in the heart of PDRs. 
  We focus our study on Spectral and Photometric Image Receiver (SPIRE) Fourier-Transform Spectrometer (FTS) fully sampled maps that allow us for the first time to study the bulk of cool/warm dust and warm molecular gas (CO) together. 
  In particular, we investigate if these populations spatially coincide, if and how the medium is structured, and if strong density and temperature gradients occur, within the limits of the spatial resolution obtained with \textit{Herschel}.}
  % Methods
   {The SPIRE FTS fully sampled maps at different wavelengths are analysed towards the northwest (NW) and the east (E) PDRs in NGC 7023.
We study the spatial and spectral energy distribution of a wealth of intermediate rotational $^{12}$CO 4$\le {\rm J_{\rm u}} \le$13 and $^{13}$CO 5$\le {\rm J_{\rm u}} \le$10 lines. 
A radiative transfer code is used to assess the gas kinetic temperature, density, and column density at different positions in the cloud.
The dust continuum emission including \textit{Spitzer}, the Photoconductor Array Camera and Spectrometer (PACS), and SPIRE photometric and the Institute for Radio Astronomy in the Millimeter Range (IRAM) telescope data is also analysed.
Using a single modified black body and a radiative transfer model, we derive the dust temperature, density, and column density.}
     % Results
   {The cloud is highly inhomogeneous, containing several irradiated dense structures.
Excited ${\rm ^{12}CO}$ and ${\rm ^{13}CO}$ lines and warm dust grains localised at the edge of the dense structures reveal high column densities of warm/cool dense matter.
Both tracers give a good agreement in the local density, column density, and physical extent, leading to the conclusion that they trace the same regions.
The derived density profiles show a steep gradient at the cloud edge reaching a maximum gas density of $10^5-10^6$ cm$^{-3}$ in the PDR NGC 7023 NW and $10^4-10^5$ cm$^{-3}$ in the PDR NGC 7023 E and a subsequent decrease inside the cloud.
Close to the PDR edges, the dust temperature (30 K and 20 K for the NW and E PDRs, respectively) is lower than the gas temperature derived from CO lines (65$-$130 K and 45$-$55 K, respectively). 
Further inside the cloud, the dust and gas temperatures are similar. 
The derived thermal pressure is about 10 times higher in NGC 7023 NW than in NGC 7023 E. 
Comparing the physical conditions to the positions of known young stellar object candidates in NGC 7023 NW, we find that protostars seem to be spatially correlated with the dense structures. }
   % Conclusion
   {Our approach combining both dust and gas delivers strong constraints on the physical conditions of the PDRs. We find dense and warm molecular gas of high column density in the PDRs. }
 \keywords{infrared: ISM - submillimeter: ISM - ISM: lines and bands - ISM: molecules  - ISM: clouds - ISM: dust, extinction}

\authorrunning{K{\"o}hler et al.}
\titlerunning{Physical structure of PDRs in NGC 7023}

\maketitle

\section{Introduction} \label{intro}

Observations of dust and molecular gas emission in photodissociation regions (PDRs) with the {\it Herschel} Space Telescope open new perspectives in our understanding of the evolution of molecular clouds. 
Several studies investigate how the physical structure in interstellar clouds is determined by the different sources of energy and the way they are injected into a cloud, such as by turbulence, gravitational contraction, stellar winds, radiative heating, and supernovae. 
Located at the interface between dense molecular clouds and diffuse regions of atomic and ionised gas, PDRs are the regions 
in which the interaction of FUV photons with matter governs the structure, dynamics, chemistry, and thermal balance of the cloud as well as the star formation associated with the cloud \citep[for a review see][]{hollenbach-tielens-1999}.
Compression of the gas can be induced by the dynamical impact from FUV radiation pressure, stellar outflows, and winds.
PDRs, which reprocess much of the radiation energy emitted by young massive stars, cool via the copious emission of IR-submm atomic and molecular lines (e.g. C$^+$ 158 $\mu$m, O$^0$ 63 and 145 $\mu$m, H${\rm _2}$ and CO). 
These lines together with dust emission are the tracers that derive the physical structure and excitation conditions. 
Most of these gas lines and dust emission fall in the wavelength range between 55 and 672 $\mu$m covered continuously for the first time by {\it Herschel}.

\begin{center}
\begin{figure*}[t]
\begin{center}
\includegraphics[width=0.95\textwidth]{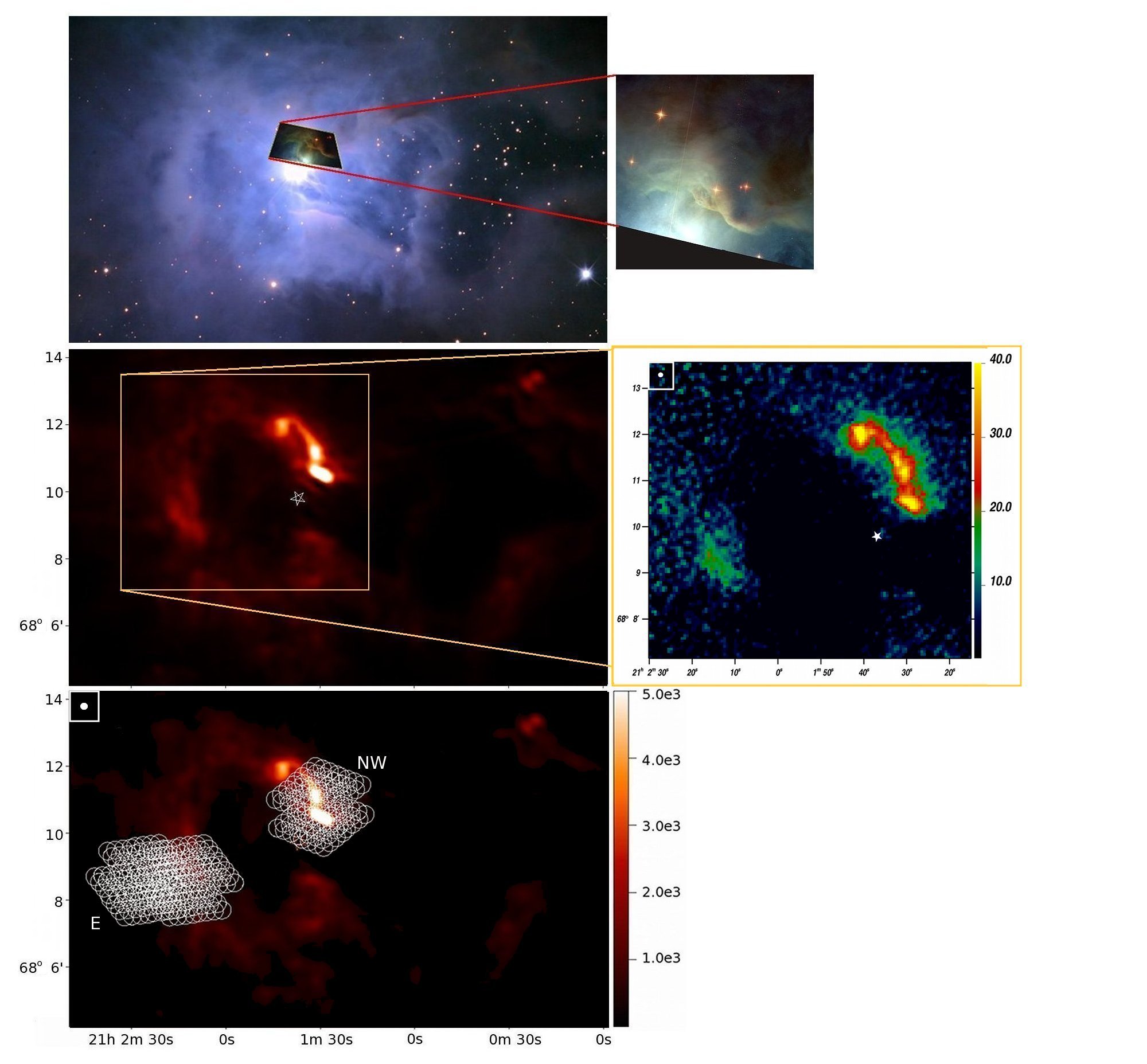}
\caption[]{
Top: visible map of NGC 7023 (from Wikisky.org) with the Hubble Space Telescope image, middle left: SPIRE photometer map at 250 $\mu$m emission revealed with the SUPREME method, the yellow box indicates the position of the image at the right, which is the IRAM-30m continuum map at 1200 $\mu$m observed at Pico Veleta using MAMBO-2, bottom: SPIRE photometer map at 250 $\mu$m emission revealed with SUPREME method with SLW jiggle positions (white circles) of {\it Herschel} SPIRE FTS observations of the northwest (NW) PDR and east (E) PDR.  The beam size of 11'' for the MAMBO-2 map and the FWHM of 11.6'' for the SPIRE photometer maps at 250 $\mu$m are shown by the white filled circles in the top left corners. All bars give the intensities in MJy/sr. For all maps, the epoch of the coordinate system is J2000. The stars indicate the position of HD 200775.}
\label{fig:full}
\end{center}
\end{figure*}
\end{center}

In this study, we focus on the PDRs in NGC 7023 (the Iris Nebula) one of the most famous and brightest nebul$\ae$ located in the Cepheus Flare region. 
The molecular cloud associated with the nebula has been shaped by the star formation process and holds the largest concentration of young stellar objects (YSOs) in the Cepheus cloud associations surveyed by {\it Spitzer} \citep{kirk-et-al-2009}.
The driving source of NGC 7023 is the massive binary system HD~200775 of B3Ve-B5 spectral type, which has been extensively studied \citep[e.g.][]{racine-1968,altamore-et-al-1980,finkenzeller-1985, pogodin-et-al-2004,witt-et-al-2006,alecian-et-al-2008} with a recently detected protoplanetary disc \citep{okamoto-et-al-2009,benistry-et-al-2013}. 
The Hipparcos parallax distance to HD~200775 is estimated to be 520$^{+180}_{-110}$ pc \citep{van-leeuwen-2007}.
A recent study from \citet{benisty-et-al-2013} reveals a distance of 320$\pm$51 pc using H$_\alpha$ line data.
In this paper, we consider an averaged distance of around 420 pc.
The stellar outflow is currently inactive, but it has generated an asymmetric east-west biconical cavity that is clearly visible at short and long wavelengths with a size of 1.5~pc~$\times$~0.8~pc (see Fig. \ref{fig:full}).
The interstellar matter is dense within the cavity wall and the star illuminates the matter in its vicinity, forming the reflection nebula.

At the edges of the cavity, three PDRs lie at $\sim$42'' northwest (herein NW), $\sim$55'' southwest (SW) and $\sim$155'' east (E) of the star (see Fig. \ref{fig:full}).
These regions, dominated by FUV photons and viewed approximately edge-on, are ideal to study the interaction of stellar FUV radiation with gas in the adjacent cloud.
The incident radiation field scales as the inverse squared distance to the illuminating star.
The FUV field intensity $G_0$ in terms of the Habing field, which corresponds to 1.6$\times$10$^{-3}$ erg cm$^{-2}$ s$^{-1}$ when integrated between 91.2 and 240~nm \citep{habing-1968}, is estimated to be $G_0$ = 2600, 1500 and 250 at the NW, SW, and E PDR fronts, respectively \citep{pilleri-et-al-2012}. 
Emission of the gas and dust associated with these PDRs have been observed extensively in the visible \citep[e.g.][]{witt-et-al-2006,berne-et-al-2008}, in the near-IR H$_2$ lines \citep{lemaire-et-al-1999}, in the mid-IR lines and very small dust particle features with ISO and {\it Spitzer} \citep[e.g.][]{fuente-et-al-2000, an-sellgren-2003, werner-et-al-2004, rapacioli-et-al-2005, fleming-et-al-2010, habart-et-al-2011} and in the radio \citep[e.g.][]{fuente-et-al-1998,gerin-et-al-1998,yuan-et-al-2013}.
NGC 7023 hosts structures at different gas densities from a very diluted atomic gas at $n_{\rm H}\sim 100$~cm$^{-3}$ in the cavity \citep{gerin-et-al-1998}, an intermediate medium at $n_{\rm H}\sim 10^4$~cm$^{-3}$ at the PDR surface \citep{pilleri-et-al-2012} to high dense filaments and clumps with $n_{\rm H}\sim 10^5-10^6$~cm$^{-3}$ \citep[e.g.][]{fuente-et-al-1996,lemaire-et-al-1996}. 
Figure \ref{fig:full} (middle, right) presents the new 1.2 mm observations with the Institute for Radio Astronomy in the Millimeter Range (IRAM) telescope, which shows evidence of several dense rather compact structures. These structures follow the edge of the cavity.

NGC~7023 has been observed in several {\it Herschel} programs providing a wealth of spatial and spectral information on gas and dust in the heart of PDRs.
\citet{abergel-et-al-2010} combined {\it Spitzer} and {\it Herschel} maps to study the spatial variations of the dust properties towards NGC 7023 E. 
The authors analyse the dust component at thermal equilibrium, which contains most of the dust mass.  
The results clearly show evidence of changes in dust properties from the diffuse interstellar medium to denser regions.
\citet{joblin-et-al-2010}, using the HIFI instrument, show that the C$^+$ emission follows the PAH emission.
\citet{okada-et-al-2013}, using both the Photoconductor Array Camera and Spectrometer (PACS) and {\it Spitzer} observations, studied the photoelectric heating efficiency in several PDRs, including NGC 7023.
The C$^+$ lines show a rich velocity structure varying with the position in the PDR, related to its complex dynamical structures (Bern\'e et al., in prep.).  
\citet{bernard-salas-et-al-2014} use the PACS observations to map the spatial distribution of the C$^+$~158~$\mu$m, O$^0$~63 and 145~$\mu$m lines towards NGC 7023 NW. 
Dense structures at the edge of the cloud can be traced via the emission of these main cooling lines, offering the opportunity to study the link between the morphology and energetics of this region.
Joblin et al. (in prep.) used the PACS scan observations of the high-J CO lines (J$_{\rm u}\ge$15) to analyse the gas excitation in the NGC 7023 NW PDR.

In this paper, we present Spectral and Photometric Image Receiver (SPIRE) Fourier-Transform Spectrometer (FTS) fully sampled maps towards NGC 7023 NW and E that allow us for the first time to study the bulk of dust and cool/warm molecular gas together.
This paper is organised as follows. 
The observations and data processing are described in Sect. \ref{sect_observation_data}, including SPIRE FTS, photometer, and IRAM Max-Planck Millimeter Bolometer Array (MAMBO-2) observations.
In Sect. \ref{sect_line_detection_spatial_SED}, we present the detected atomic and molecular lines.
In Sect. \ref{spatial-7023NW}, we present the spatial distribution of the CO lines and the dust emission.
The cooling curves of CO molecules are presented in Sect. \ref{SED-7023NW}.
In order to assess the physical conditions in which the observed CO lines arise, we analyse the integrated line intensities using a non-local thermodynamic equilibrium (NLTE) simple radiative transfer code in Sect. \ref{SED-7023NW}.
In Sect. \ref{dust}, the dust emission observed with {\it Spitzer} and {\it Herschel} is analysed as an independent tracer. 
The results from the dust and molecular and atomic gas lines are compared and discussed in Sect. \ref{dust-gas}.
In Sect. \ref{star}, we discuss the link between the physical conditions in the molecular cloud and the star formation. Finally, the conclusions are presented in the last section.

\begin{center}
\begin{figure*}[t]
\includegraphics[width=0.99\textwidth]{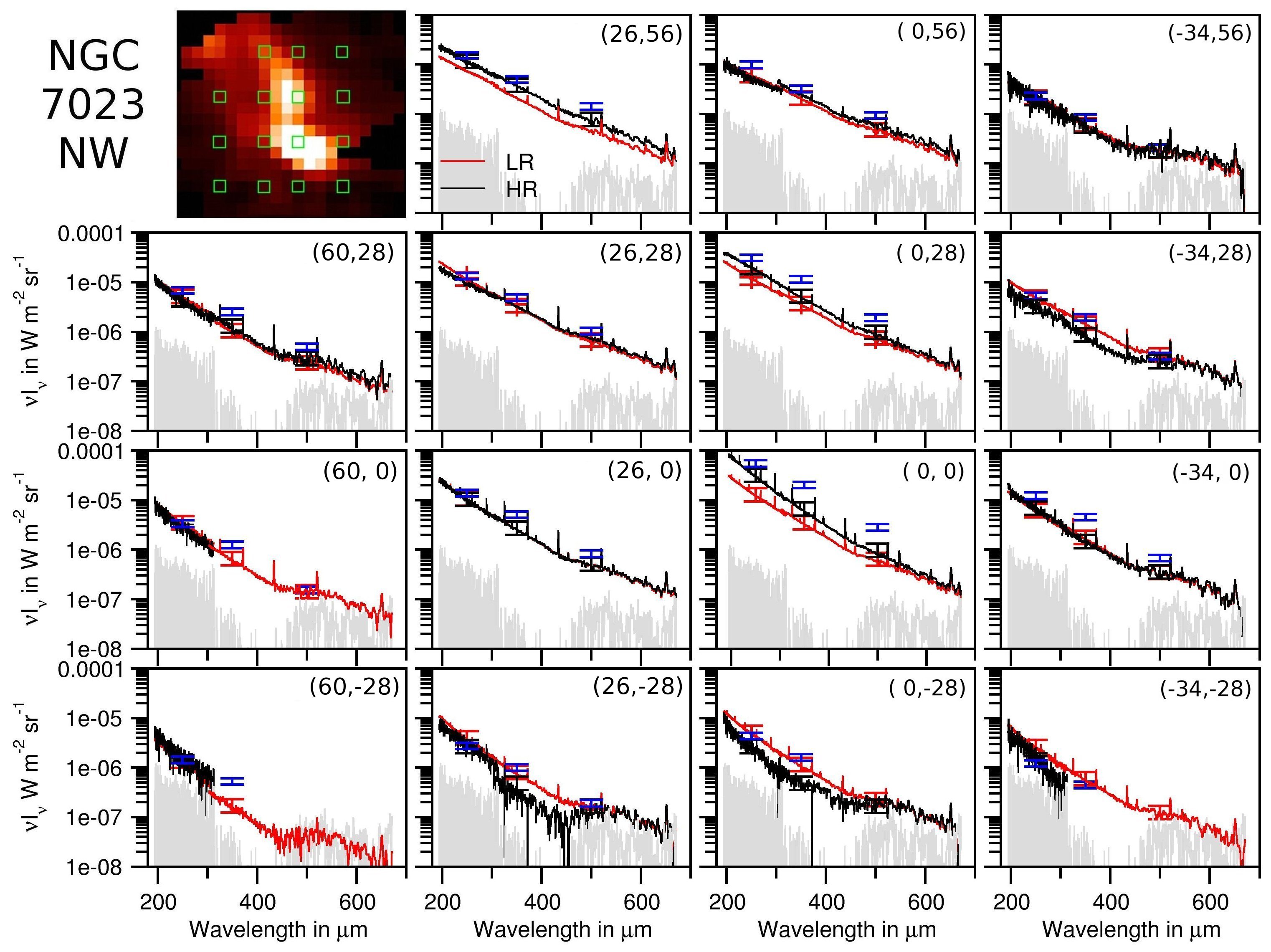}
\caption[]{
The obtained FTS spectra with the HR cube (black curves) and photometer data (blue signs) at different positions in NGC 7023 NW indicated as green squares in the map. The black signs are the HR FTS data integrated over the photometric filters using the extended source calibration. The red curves indicate the LR spectrum and the red signs the LR FTS data integrated over the photometric filters using the extended source calibration. The grey curves present the background spectrum. Some positions are outside the observation area for SLW. Calibration errors of 15\% for the photometer and 30\% for the integrated FTS data are included. The offsets in arcsec from the position of the brightest spectrum are included.}
\label{fig:speclineN}
\end{figure*}
\end{center}

\begin{center}
\begin{figure*}[t]
\includegraphics[width=0.99\textwidth]{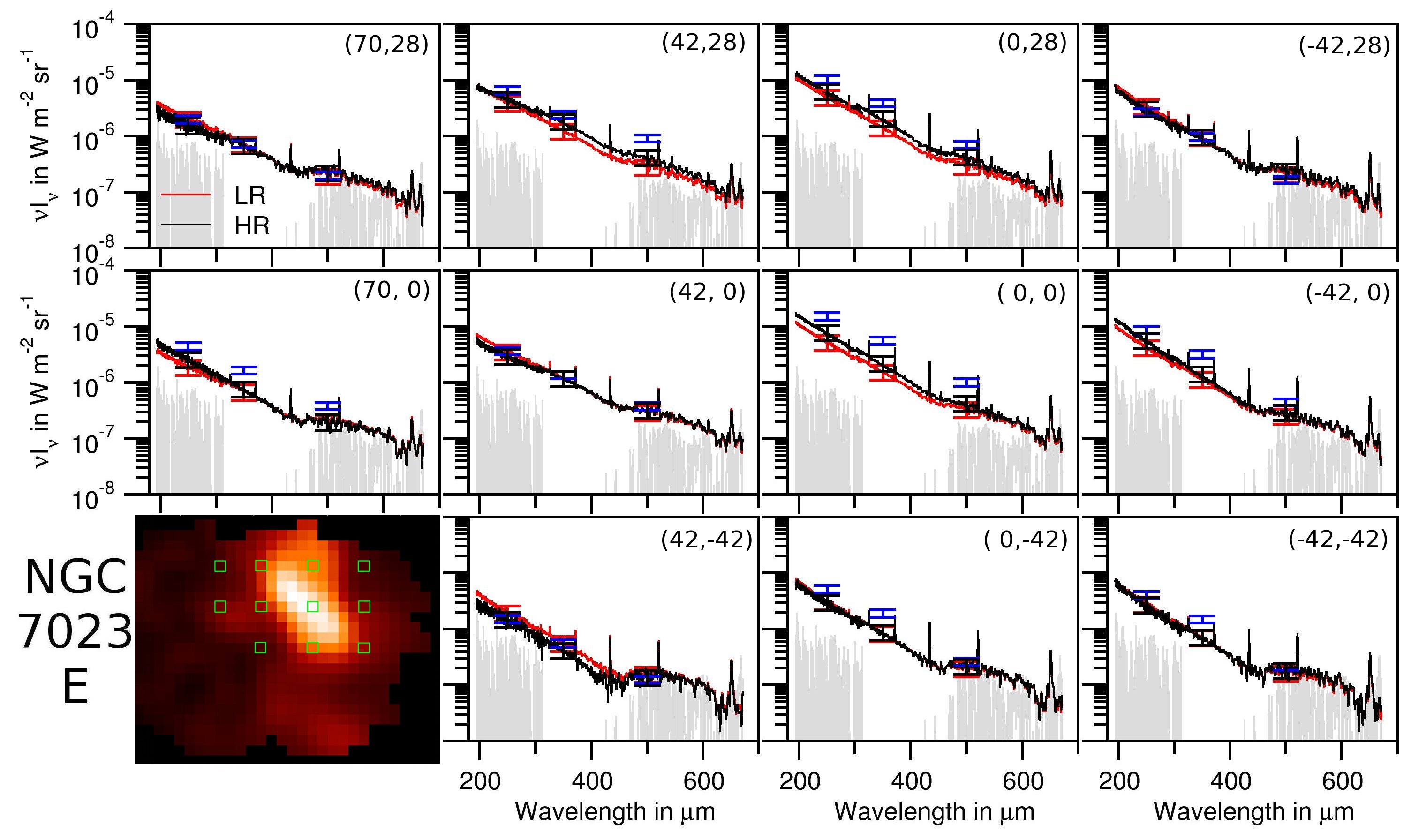}
\caption[]{Same as in Fig. \ref{fig:speclineN} for NGC 7023 E.}
\label{fig:speclineE}
\end{figure*}
\end{center}

%%%%%%%%%%%%%%%%%%%%%%%%%%%%%%%%%%%%%%%%%%%%%%%%%%%%%%%%%%%%%%%%%%%%

\section{Observation and data processing}
\label{sect_observation_data}

\subsection {SPIRE observations}

NGC 7023 was observed in photometry and spectroscopy with the {\it Herschel} SPIRE instrument, as part of the {Evolution of Interstellar Dust} key program \citep{abergel-et-al-2010}.
Large maps of 8' x 8' were performed at three wavelengths, 250, 350, and 500 $\mu$m \citep{abergel-et-al-2010}.  
For spectroscopy, three pointings, one on the PDR in the NW direction ({\it Herschel} ID 1342198923) and two on the PDR in the E direction ({\it Herschel} ID 1342201204 and 1342201205) were carried out in the high-resolution full-sampling mode of the SPIRE FTS (see Fig. \ref{fig:full}).
The wavelength range extends from 194 to 671 $\mu$m where the spectrometer long wavelength (SLW) covers the wavelength range of 303$-$671 $\mu$m and the spectrometer short wavelength (SSW) covers the wavelength range of 194$-$313 $\mu$m. 
The FTS beam size and shape vary with wavelength and cannot be characterised by a simple Gaussian response. 
This is expected from the multi-moded feedhorns used for the spectrometer arrays \citep{makiwa-et-al-2013}.
The FWHM beam-widths using Gaussian fits varies from 16.8'' to 42'' from small to long wavelengths. For further details on the beam size see \citet{makiwa-et-al-2013}.
The total observation time was 4883 s for each of the three positions. The observations were carried out on the 22nd of June 2010 for NGC 7023 NW and the 26th of July 2010 for NGC 7023 E.

The data processing was performed using HIPE 8.
The FTS calibration and data reduction procedures are explained in detail in the article by \citet{swinyard-et-al-2010}.
In the following, we present the gridding and super-resolution method SUPREME\footnote{http://www.ias.u-psud.fr/supreme/} used for FTS data, the FTS-photometer cross calibration, and the line fitting.
The reduced data cubes are available on the Herschel Idoc Database (HESIOD) webpage\footnote{http://idoc-herschel.ias.u-psud.fr/sitools/client-user/}.

\subsubsection{Gridding and super-resolution}

The {\it Herschel} FTS data are irregularly distributed over the observed area and we therefore carry out a gridding that computes the spectra at positions on a regular square grid. 
The grid points are determined by a bilinear interpolation weighting the data points by their distance to the grid points.
This avoids the simplicity of the Naive or the Nearest Neighbour gridding methods and the smoothing of the Heterodyne Instrument for the Far-Infrared (HIFI) gridder, which are implemented in the Herschel Interactive Processing Environment (HIPE) pipeline.
The super-resolution method SUPREME is applied to the FTS data (Ayasso et al., in prep.).
It is based on a Bayesian approach and uses an instrument model to obtain cubes with a higher spatial resolution than the nominal resolution of the detector. 
Corrections of the point spread functions (PSFs) were made using a beam profile model provided by the University of Lethbridge \citep{makiwa-et-al-2013}.
This method provides two cubes, one with high resolution for each wavelength and another one with a controlled unique equivalent PSF.
In Figs. \ref{fig:full}, \ref{fig:linedistrnw}, \ref{fig:250mic}, and \ref{fig:linedistre}, we give a FWHM using a Gaussian fit with the same bandwidth of the equivalent beam. 
It should however be noted that the PSF is non-Gaussian and a FWHM does not account for the shape of the beam. 
The gain in resolution with the SUPREME method, for example, is from a FWHM of 16.6'' to 11.9'' at 200 $\mu$m, from 32.6'' to 19.0'' at 400 $\mu$m and from 36.5'' to 24.1'' at 600 $\mu$m.
For further details and a description of the PSF see Ayasso et al. (in prep).

In Figs. \ref{fig:speclineN} and \ref{fig:speclineE}, we present the observed spectrum at different positions for NGC 7023 NW and NGC 7023 E, respectively.
The black and red curves show the spectrum extracted from the cubes at high spatial resolution (herein HR) and controlled unique equivalent PSF (i.e. low spatial resolution, herein LR), respectively.
In the HR case, the spatial resolution changes with wavelength, while it is constant for the LR case. 
The HR and LR spectra are equivalent at the longest wavelengths, while at short wavelengths differences in the spectral shapes and absolute intensities occur.
For the NW PDR the intensity in the LR spectrum is lower than in the HR spectrum because of dilution effects.
The observed emission comes from a compact, semi-extended source that is not extended over the largest beam size. 
For NGC 7023 E, we note that the differences in the HR and LR spectrum are smaller.
The emission is spatially more extended than in NGC 7023 NW.
In the following, we consider both the HR and LR to use advantages of both: (1) high resolution at small wavelengths (HR) and (2) consistent beam size (LR) to analyse the full spectrum and the complete CO ladders.
All maps are shown in HR. For all analyses of spectral energy distributions of gas lines we use LR data.

\subsubsection{The FTS-Photometer cross calibration}

The flux calibration for the FTS observations uses a relative spectral response function (RSRF) based on the telescope model emission. By default it is assumed that the observed source is uniformly extended over the entire beam, and the calibration uncertainty due to flat fielding across the array is 10$-$15\% 
\citep{observermanual}. %(SPIRE Observer's Manual, 2011). 
However, any source morphology would affect the calibration \citep{wu-et-al-2013} and so we take a conservative approach and estimate the calibration accuracy of our maps to be 15-30\%, depending on position in the bands \citep{griffin-et-al-2010, swinyard-et-al-2014}.

In Figs. \ref{fig:speclineN} and \ref{fig:speclineE}, we compare the spectrometer to the photometer data.
For cross calibrations, we use the photometer data processed with the SUPREME method \citep{ayasso-et-al-2012}.
The calibration uncertainty for the photometer is conservatively estimated as $\pm$7\% for point sources and 10$-$15\% for extended sources (http://Herschel.esac.esa.int/), and we include an error of 15\% on the presented data.
We integrate the FTS data (HR and LR) over the photometric filters assuming the RSRF of the photometer for extended sources. 
The differences in the FTS data integrated in the photometer filter, assuming an extended source, are smaller than 10\% compared to the FTS data integrated in the photometer filter, assuming a point source (one exception with differences around 40\%).
The comparison between the FTS and photometer data shows that the intensity values can deviate up to 70\%, which occurs at positions of bright emission.
At these positions, the HR case gives a better agreement with the photometer than the LR case.
Where the signal-to-noise ratio is low, the FTS data show a bump at long wavelengths.
This results from background subtraction problems \citep{swinyard-et-al-2014}.
We include the background spectrum (grey curves) in Figs. \ref{fig:speclineN} and \ref{fig:speclineE}.

\begin{center}
\begin{figure*}[t]
\includegraphics[width=0.99\textwidth]{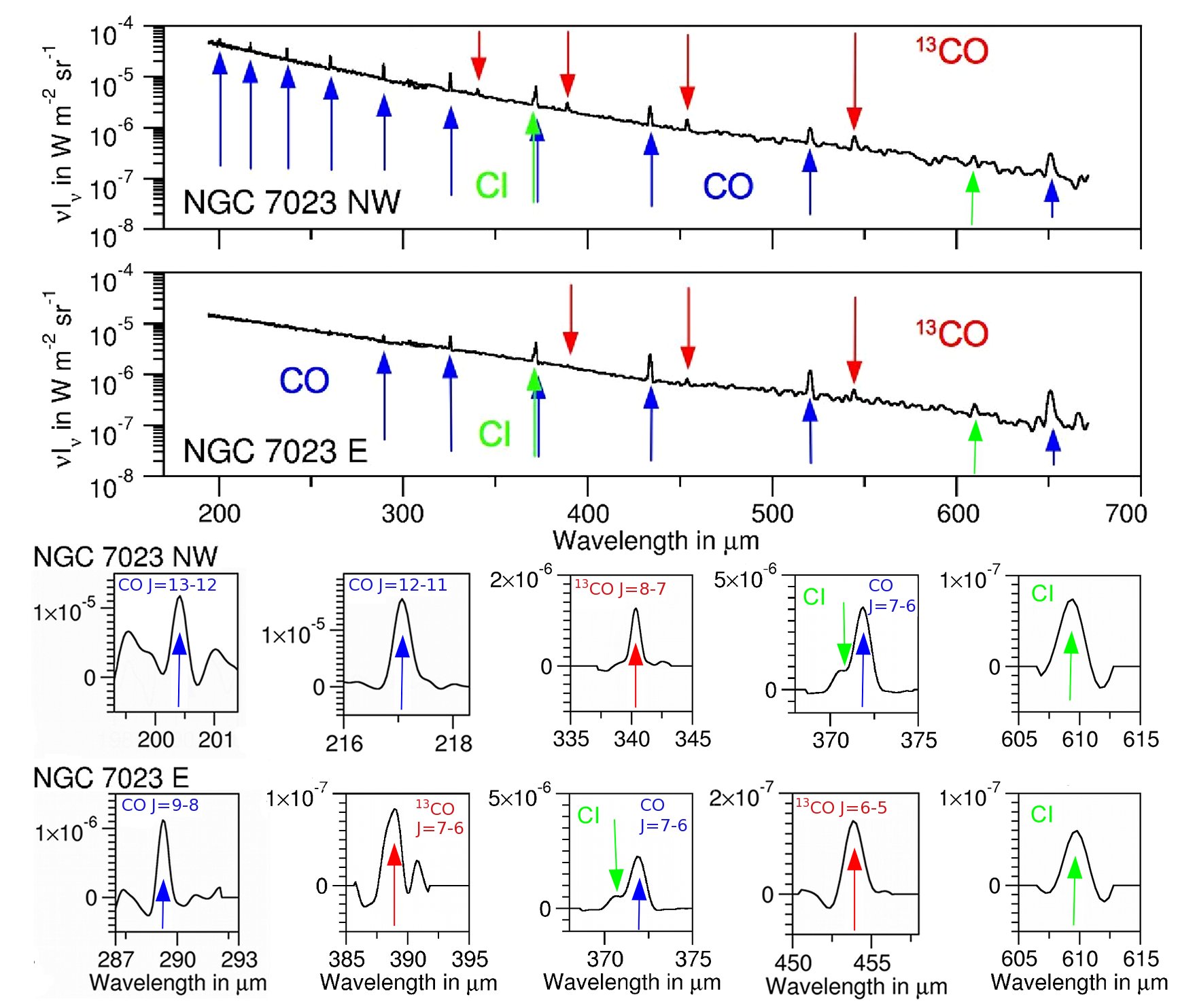}
\caption[]{Spectrum at the position of brightest emission for NGC 7023 NW and NGC 7023 E (top), showing the HR spectrum. A zoomed-in spectrum is shown for the weaker lines {where we subtracted the continuum emission}. For NGC 7023 NW and E, these positions of brightest emission are called position A in the following (see Fig. \ref{fig:250mic}).}
\label{fig:brightspec}
\end{figure*}
\end{center}

\begin{center}
\begin{figure*}[t]
\includegraphics[width=0.7\textwidth]{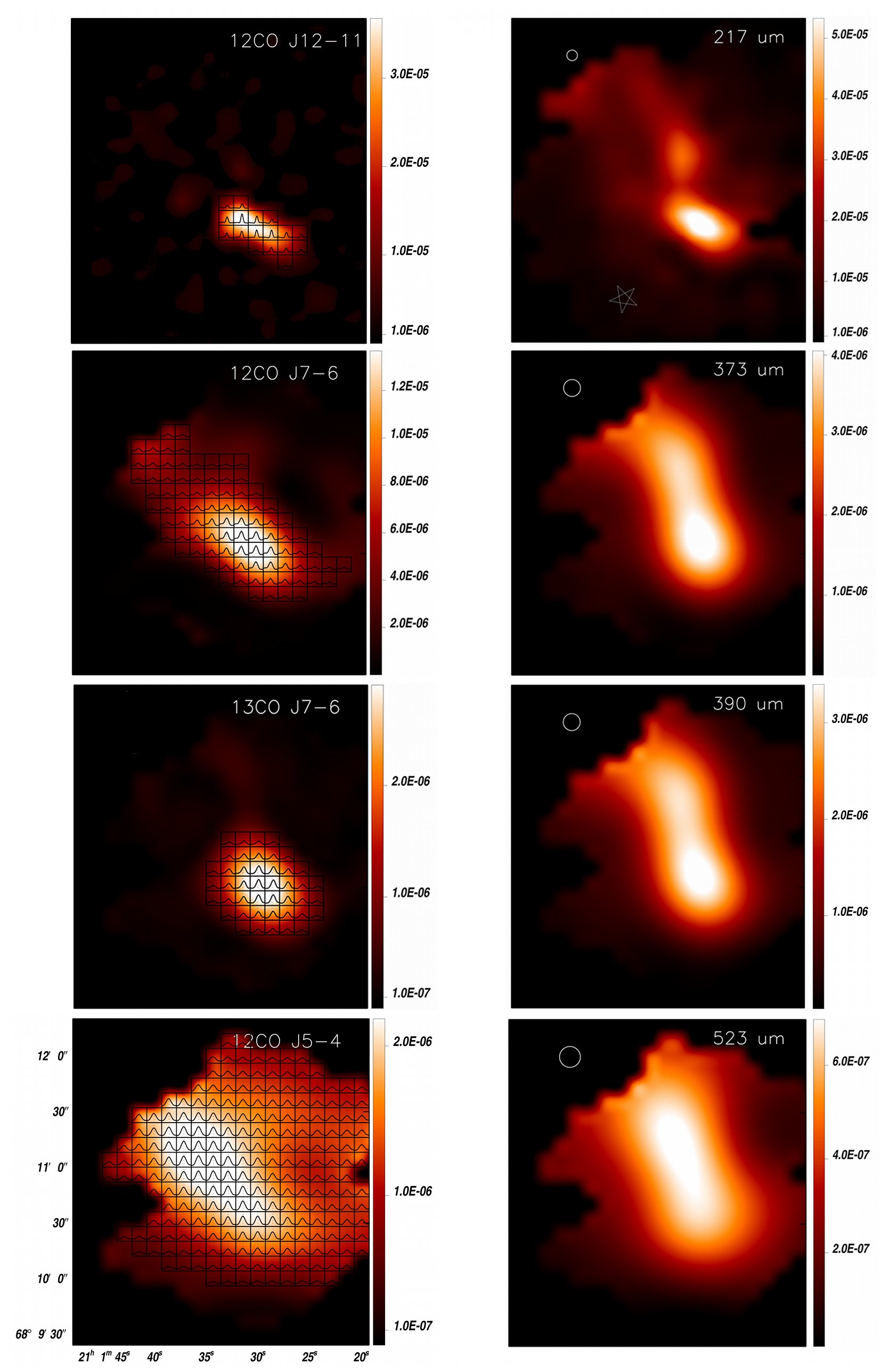}
\caption[]{Left column: Spatial distribution of ${\rm ^{12}CO}$ (J=12$-$11), ${\rm ^{12}CO}$ (J=7$-$6), ${\rm ^{13}CO}$ (J=7$-$6) and ${\rm ^{12}CO}$ (J=5$-$4) for NGC 7023 NW. The bars give intensities in ${\rm erg~s^{-1} cm^{-2} sr^{-1}}$.
Right column: {Spatial distribution of dust emission} at the wavelength close to the line emission.
The bars give ${\rm \nu I_{\nu}}$ in ${\rm W m^{-2} sr^{-1}}$. All images are presented with the same scale, same field, and same centre. We include the FWHM of the beam in the dust emission maps: 11.6'', 18.6'', 18.7'', 21.3'' from top to bottom.}
\label{fig:linedistrnw}
\end{figure*}
\end{center}

\begin{figure*}[t]
\begin{center}
\includegraphics[width=0.99\textwidth]{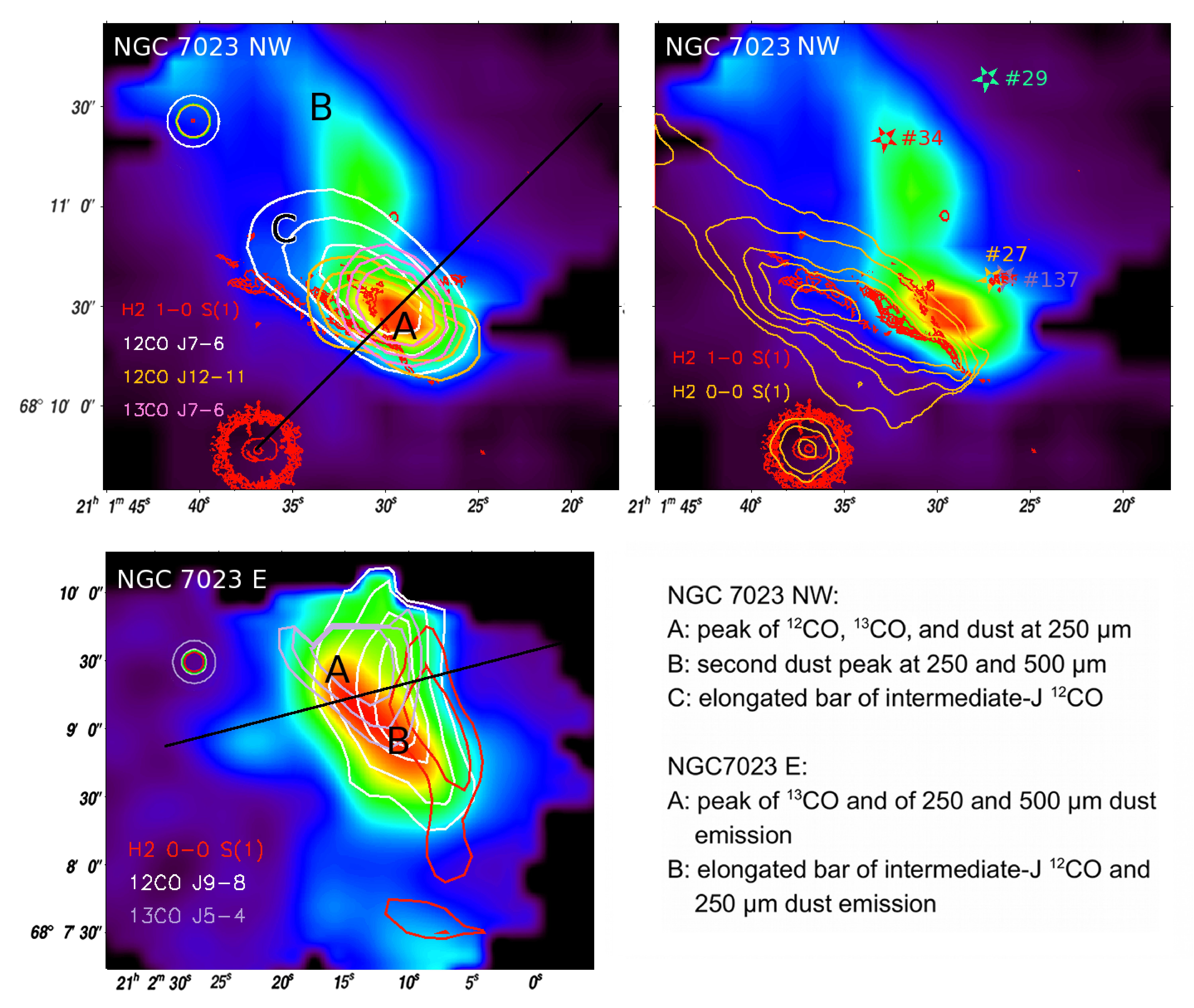}
\caption[]{
Top left: {FTS map at 250 $\mu$m of the PDR NGC 7023 NW} with contours of 
${\rm ^{12}CO}$ J=7$-$6 (white), 
(maximum at $1.5 \times 10^{-5}$ ${\rm erg~cm^{-2} s^{-1} sr^{-1}}$ with contour lines at 0.4, 0.6, 0.8 times the maximum), 
${\rm ^{12}CO}$ J=12$-$11 (orange) 
(maximum at $4.0 \times 10^{-5}$ ${\rm erg~cm^{-2} s^{-1} sr^{-1}}$ with contour lines at 0.2, 0.4, 0.6 times the maximum), 
and ${\rm ^{13}CO}$ J=7$-$6 (pink) 
(maximum at $3.2 \times 10^{-6}$ ${\rm erg~cm^{-2} s^{-1} sr^{-1}}$ with contour lines at 0.4, 0.6, 0.8 times the maximum) 
and ${\rm H_2}$ 1$-$0 S(1) (red) from \citet{lemaire-et-al-1996}
(maximum at  $3.7 \times 10^{-4}$ ${\rm erg~cm^{-2} s^{-1} sr^{-1}}$ with contours at 0.27, 0.41, 0.55, 0.69 times the maximum). 
Top right: {FTS map at 250 $\mu$m of the PDR NGC 7023 NW} with contours of 
${\rm H_2}$ 0$-$0 S(1) (orange) from {\it Spitzer} observations
(maximum at $1.9 \times 10^{-4}$ ${\rm erg~cm^{-2} s^{-1} sr^{-1}}$ with contour lines at 0.26, 0.36, 0.52, 0.62, 0.83 times the maximum)
and ${\rm H_2}$ 1$-$0 S(1) (red) 
(maximum at  $3.7 \times 10^{-4}$ ${\rm erg~cm^{-2} s^{-1} sr^{-1}}$ with contours at 0.27, 0.41, 0.55, 0.69 times the maximum). 
Bottom: FTS map of the PDR NGC 7023 E, 250 $\mu$m map with contours of 
${\rm ^{12}CO}$ J=9$-$8 (white) 
(maximum at $3.2 \times 10^{-6}$ ${\rm erg~cm^{-2} s^{-1} sr^{-1}}$ with contour lines at 0.4, 0.6, 0.8 times the maximum), 
${\rm ^{13}CO}$ J=5$-$4 (violet) 
(maximum at $5.6 \times 10^{-7}$ ${\rm erg~cm^{-2} s^{-1} sr^{-1}}$ with contour lines at 0.7, 0.8, 0.9 times the maximum) 
and ${\rm H_2}$ 0$-$0 S(1) (red) from {\it Spitzer} observations
(maximum at $1.3 \times 10^{-4}$ ${\rm erg~cm^{-2} s^{-1} sr^{-1}}$ with contour lines at 0.45, 0.66 times the maximum). 
Note that the FTS map at long wavelengths (SLW) is slightly smaller than the shown map, so that the ${\rm ^{13}CO}$ J=5$-$4 contour lines are cut off in the north.
The letters A, B, and C indicate positions of interest as explained. We include the detected YSOs by \citet{kirk-et-al-2009} as stars in the top right figure. 
The FWHM of the beam for the different CO and the beam size for H$_2$ lines are presented. Maps: 12.2'' (green). NGC 7023 NW: 11.6'' for $^{12}$CO J=12$-$11 (orange),  18.6'' for $^{12}$CO J=7$-$6 (white), 18.7'' for $^{13}$CO J=7$-$6 (white) and 1'' for H$_2$ 1$-$0 S(1) (red). NGC 7023 E: 13.0'' for $^{12}$CO J=9$-$8 (white), 22.7'' for $^{13}$CO J=5$-$4 (violet) and 11'' for H$_2$ 0$-$0 S(1) (red).}
\label{fig:250mic}
\end{center}
\end{figure*}

\subsubsection{Line fitting method} \label{sec:linefit}

The observed unapodized spectra have a resolution of 0.04 cm$^{-1}$ and the apodized spectra have a resolution of 0.07 cm$^{-1}$.
In any case, the detected lines are not resolved with FTS and the instrumental line width ($\Delta \rm{v}=230-800~{\rm km~s}^{-1}$) dominates.

We fit the lines by determining a linear fit to the continuum of the apodized spectrum in the continuum adjacent to each line. 
We then subtract this continuum and fit the remaining line with a Gaussian.
We then calculate the integrated line intensities by integrating over the Gaussian fit.
We assume a conservative total error of 30\% for the integrated line intensities, which includes the calibration uncertainties plus the line fitting errors.

\begin{table*}[t]
\caption{The $^{12}$CO and $^{13}$CO lines observed in the spectra of NGC 7023 NW and NGC 7023 E. The wavelength ($\lambda$), wavenumber (WN), and $E_u$ are presented in columns 3, 4, and 5. The last five columns give the low spatial resolution (LR) integrated line intensities, $I_{\nu}$, in erg/s/cm$^2$/sr of the positions A, B, C in NGC 7023 NW and A, B in NGC 7023 E (shown in Fig. \ref{fig:250mic}).}
\begin{tabular}{lcccc|ccc|cc}
\hline
     &   Transition  &   $\lambda$  &  WN   & $E_u$                       	& \multicolumn{3}{c|}{NGC7023 NW}         & \multicolumn{2}{c}{NGC 7023 E} \\
     &		           &   [$\mu$m]    & [cm$^{-1}$] & [K] &                       A 		& B 		& C        & A         & B   \\
\hline
$^{12}$CO \T &   J=4$-$3    &  650.3    &   15.4 & 55.3			&   6.9e-07    &     8.2e-07    &     7.3e-07    &     1.1e-06    &  1.4e-06\\
$^{12}$CO &   J=5$-$4  &  520.2    &   19.2 &  83.0  			&   3.3e-06    &   3.5e-06    &     5.5e-06    &     4.7e-06    &   3.5e-06   \\
$^{12}$CO   &  J=6$-$5    &     433.6    &   23.1 & 116.1		&   3.3e-06    &   3.5e-06    &     5.5e-06    &     4.7e-06    &   3.5e-06 \\
$^{12}$CO    &  J=7$-$6    &    371.7    &  26.9 &  154.9		&   8.0e-06    &   5.3e-06    &     7.9e-06    &     5.6e-06    &   4.8e-06    \\
$^{12}$CO   &    J=8$-$7    &    325.2    &   30.8 & 199.1		&   1.1e-05    &   5.3e-06    &   9.1e-06    &   4.7e-06    &   4.0e-06        \\
$^{12}$CO   &   J=9$-$8    &    289.1    &   34.6 &  248.9 		&   1.1e-05    &   3.8e-06    &   7.2e-06    &    2.5e-06    &   1.7e-06   \\
$^{12}$CO   &   J=10$-$9    &    260.2    &   38.4 &  304.1		&   1.0e-05    &   2.1e-06    &   5.9e-06    &   6.0e-07    &   -    \\
$^{12}$CO   &   J=11$-$10    &    236.6    &   42.3 &  365.0 	&   8.2e-06    &   -    &   3.4e-06    &   -    &   -    \\
$^{12}$CO   &    J=12$-$11    &   216.9    &   46.1 &  431.3 	&   8.6e-06    &   -    &   -    &   -    &   -   \\
$^{12}$CO   \B&   J=13$-$11    &    200.3    &   49.9 &  503.1  	&   5.0e-06    &   -    &   -    &   -    &   -  \\
\hline
$^{13}$CO  \T&   J=5$-$4    &    544.2    &   18.4 &  79.33		&   7.0e-07    &   5.1e-07    &  5.0e-07    &     4.1e-07   &     2.4e-07   \\
$^{13}$CO  &   J=6$-$5    &    453.5    &   22.1 &  111.1		&   1.0e-06    &   5.5e-07    &  6.8e-07    &     3.5e-07   &     2.8e-07   \\
$^{13}$CO  &   J=7$-$6    &    388.7    &   25.7 &  148.1		&   1.6e-06    &   3.2e-07    &   4.9e-07    &   -   &   -    \\
$^{13}$CO  &   J=8$-$7    &    340.2    &   29.4 &   190.4		&   1.4e-06    &   -    &   -    &   -   &   -  \\
$^{13}$CO  &   J=9$-$8    &    302.4    &   33.1 &   237.9		&   1.6e-06    &   -    &   -    &   -   &   -    \\
$^{13}$CO  \B&   J=10$-$9    &    272.2   &   36.7  &   290.8	&   8.7e-07    &   -    &   -    &   -   &   -    \\
\hline
\end{tabular}
\label{tab:lines1}
\end{table*}

\subsection{The IRAM-30m/MAMBO-2 observations at 1200\,$\mu$m}

In order to obtain additional spatial and spectral information on the dust emission, we include a 1.2 mm high angular resolution map.
The millimeter continuum emission map of the region, first presented here,  was obtained using the 117--channel MAMBO-2 \citep[][]{kreysa-et-al-1998} at the \textit{IRAM-30m} telescope (Pico Veleta, Spain).
MAMBO-2 has a half-power spectral bandwidth from 210 to 290\,GHz, with the effective bandwidth centre at $\sim$250\, GHz ($\sim$1200\,$\mu$m). 
The effective beam FWHM is 10.7$''$ and the instantaneous (under-sampled) field of view is  $\sim$4$'$$\times$4$'$. 
A fast on-the-fly mapping mode  \citep{teyssier-sievers-1999} was used to map NGC\,7023 NW and NGC\,7023 E over a $\sim$10$'$$\times$10$'$ region. 
Observations were carried out during  2005 winter pool.
The total integration time was two hours, achieving a rms noise of $\sim$5\,mJy/beam.
Sky noise subtraction and  data analysis were carried out with the \texttt{MOPSIC} software \citep{zylka-1998}. 
The absolute pointing accuracy and the overall  calibration uncertainty are estimated to be  $\sim$2$''-$3$''$ and $\sim$20\%, respectively.
The resulting $\sim$1200\,$\mu$m emission map is shown in Figure~\ref{fig:full}.

%%%%%%%%%%%%%%%%%%%%%%%%%%%%%%%%%%%%%%%%%%%%%%%%%%%%%%%%%%%%%%%%%%%%

\section{Line detection}
\label{sect_line_detection_spatial_SED}

In this section, we present the spectra (Fig. \ref{fig:brightspec}) at the position of strongest warm dust and high-J CO emission and the detected lines (Tab. \ref{tab:lines1}) at different positions for NGC 7023 NW and NGC 7023 E.
The comparison of the spectra shows that more lines are detected in NGC 7023 NW than in NGC 7023 E.
In the following, we divide the  ${\rm ^{12}CO}$ lines observed in the FTS wavelength range into three groups: low-J ${\rm ^{12}CO}$ for J$_{\rm u}\le$5, intermediate-J ${\rm ^{12}CO}$ from J=6$-$5 to J=9$-$8, and high-J ${\rm ^{12}CO}$ for J$_{\rm u}\ge$10.

In NGC 7023 NW, the entire ladders of ${\rm ^{12}CO}$ from J$_{\rm up}=4$ to 13 and of ${\rm ^{13}CO}$ from J$_{\rm up}=5$ to 11 are observed.
The J$_{\rm up}=4$ to 13 rotational levels of CO lie between 55 to 503~K above ground and have critical densities of about 10$^5$ to 10$^7$ cm$^{-3}$ for collisions with H$_2$ at $T$=100~K \citep{yang-et-al-2010}. 
If collisions dominate their excitation, their detection constitute an excellent tracer of both warm and dense gas.
The gas density in molecular condensations could be large enough to approximately thermalise the CO transitions.
Their line intensities will thus depend on the kinetic temperature and molecule column density in the optically thin case. 
For optically thick lines, the kinetic temperature is the main parameter determining the intensity.

For NGC 7023 E, we detect ${\rm ^{12}CO}$ from J$_{\rm up}=4$ to 10 and ${\rm ^{13}CO}$ from J$_{\rm up}=5$ to 6 at the position of brightest 250 $\mu$m dust emission. 
The non-detection of the higher-J ${\rm ^{12}CO}$ or ${\rm ^{13}CO}$ lines indicates that this PDR is cooler and/or less dense than NGC 7023 NW.

In both PDRs, we detect the atomic carbon C$^0$ fine structure lines $^3$P$_1-^3$P$_0$ and $^3$P$_2-^3$P$_1$ at 609 and 370 $\mu$m, respectively.
The levels lie 23.6 and 62.4~K above ground and have low critical densities of about $3 \times 10^2$ to $2 \times 10^3$ cm$^{-3}$, respectively, for collisions with H at $T$=100~K \citep{launay-roueff-1977}. 
As the critical densities are low, the gas density will be large enough in the translucent region to thermalise the  C$^0$ transitions.
Their line intensity will depend on the temperature and the column density rather than the density.
The C$^0$ at 609 $\mu$m should be especially detectable further inside the PDR.
The C$^0$~370 to 609~$\mu$m intensity ratio depends on the temperature up to 100~K. 
This ratio is found to be larger (by a factor of 2.5) in NGC 7023 NW than in NGC 7023 E.

The  C$^0$~370~$\mu$m line is blended with the $^{12}$CO J=7$-$6 rotational transition. 
The $^{12}$CO to C$^0$ 370~$\mu$m intensity ratio depends on the temperature and column density of CO and  C$^0$ in the optically thin case. 
In both PDRs, the CO line is stronger than the C$^0$ line. 
In NGC 7023 NW, the CO to C$^0$ ratio is about 7 at the brightest position and decreases to 4 at the edge.
In NGC 7023 E, the CO to C$^0$ ratio is about 4 at the peak and decreases to 1.4 at the edge.

The N$^+$ fine structure line $^3$P$_1-^3$P$_0$ at 205 $\mu$m is not detected in the observed field of view.
With an ionisation potential of 14.5 eV, the N$^+$ line is expected to come from the ionised region.
The level lie 70.1~K above ground and since the critical density is very low ($<$100~cm$^{-3}$), the gas density in the ionised regions should be large enough to thermalise that transition.
The non-detection indicates that N$^+$ cannot be found in the cavity or that the column density of ionised gas in the cavity of the nebula is small and therefore the gas inside the cavity is mostly neutral.
This is expected considering the spectral type of the central stars (B3Ve-B5).
The other N$^+$ line $^3$P$_2$-$^3$P$_1$ at 122~$\mu$m observable with PACS is not detected either \citep{bernard-salas-et-al-2014}. 

%%%%%%%%%%%%%%%%%%%%%%%%%%%%%%%%%%%%%%%%%%%%%%%%%%%%%%%%%%%%%%%%%%%

\section{Spatial distribution of the CO lines and the underlying dust continuum emission} \label{spatial-7023NW}

From the observed FTS data cube, we directly obtain information about the spatial distribution of the gas lines and dust continuum emission, which we present in this section for NGC 7023 NW and NGC 7023 E.
We compare our results with {\it Spitzer} and ground-based observations.

\subsection{NGC 7023 NW}

Figure \ref{fig:linedistrnw} shows maps in different lines of ${\rm ^{12}CO}$ and ${\rm ^{13}CO}$ and the maps of the underlaying dust continuum emission. 
In the line maps, after the subtraction of the dust continuum, the line spectrum is overplotted for each pixel in which the lines have an intensity larger than 20\% of the maximal line intensity occurring in the map. 
Figure \ref{fig:250mic} presents the FTS maps of the dust emission at 250~$\mu$m with contours of the  ${\rm ^{12}CO}$~J=12$-$11, ${\rm ^{12}CO}$~J=7$-$6, and ${\rm ^{13}CO}$ J=6$-$5 line emission.

The maps show, firstly, a spatial correlation between the ${\rm ^{12}CO}$~J=12$-$11, ${\rm ^{13}CO}$~J=6$-$5 and dust emission at shorter wavelengths (around 250 $\mu$m) in a small area located in the southern region. 
This specific region, with its bright edge observed in the visible (see Fig. \ref{fig:full}), may present an area with a high column density of warm and dense gas.
From the map lines of the highest excited ${\rm ^{12}CO}$ that is observed, we derive the projected width and extent, which are the sizes perpendicular to the line of sight. 
The width is the smallest and the extent is the largest size of the line emission.
The projected width of the ${\rm ^{12}CO}$ J=12$-$11 emission is about $\le$15$-$20'' (or $\le$0.03$-$0.04~pc), which is marginally resolved by the FTS (see Fig. \ref{fig:linedistrnw}, top left).
The projected length of the emission is about 30$-$50'' (0.06$-$0.1~pc). 
The size of the emission is defined as the region where the CO line is detected with a signal-to-noise $>$4.
Secondly, another dust emission peak, which is brightest at wavelengths around 500 $\mu$m, is observed further north ($\alpha_{\rm 2000}$= 21h 1m 28s, $\delta_{\rm 2000}$= 68$^{\circ}$ 11' 30''). 
In this northern region, no {\rm intermediate-J and high-J} ${\rm ^{12}CO}$ and  ${\rm ^{13}CO}$ lines are detected. 
The gas in this region further away from the star might be colder.
Thirdly, the emission peak of the ${\rm ^{12}CO}$ J=5$-$4 and J=7$-$6 lines is seen in an elongated bar from southwest to northeast.
This bar follows the illuminated edge of the cloud as seen in the visible (Fig. \ref{fig:full}).

\begin{figure*}[ht]
\begin{center}
\includegraphics[width=0.99\textwidth]{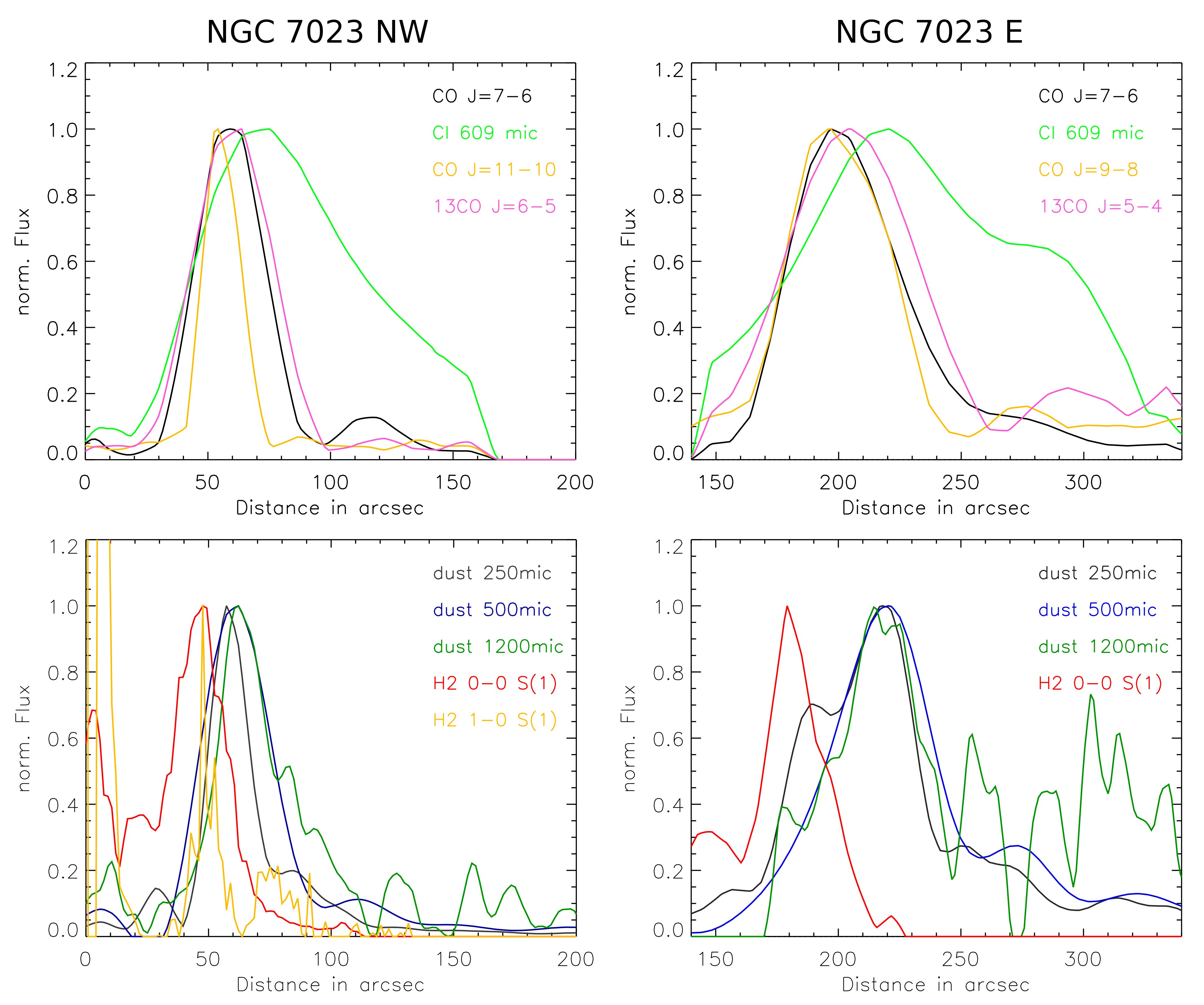}
\caption[]{Emission profiles for CO and C$^0$ (top) and dust and H$_{2}$ (bottom) for NGC 7023 NW (left) and NGC 7023 E (right) along the cuts shown in Fig. \ref{fig:250mic}, as a function of the distance from the star in arcsec. We use the HR data for this plot, so that the beam size is not constant with wavelength, which affects the width of the emission profiles. Angular resolutions are $\sim$1'' for H$_2$ 1$-$0 S(1), $\sim$10'' for H$_2$ 0$-$0 S(1), between $\sim\le$17$-$40'' for CO lines and dust emission, $\sim$40'' for C$^0$ 609 $\mu$m, and 11'' for the dust emission at 1.2 mm (IRAM-30m data).}
\label{fig:position}
\end{center}
\end{figure*}

In Fig. \ref{fig:position} we plot the spatial profile of the gas lines and dust emission for the HR data along the cuts in the direction of the illuminating star (black line in Fig. \ref{fig:250mic}).
The SPIRE resolution does not allow us to resolve in detail the spatial stratification expected from the attenuation of the stellar FUV radiation field into the cloud.
However, we are able to determine, that the CO and the dust emissions peak at about the same position at around 50$-$60'' distance from the star.
The dust and gas emission profiles reveal different widths due to beam effects and radiative and excitation processes.
We include the profile of the C$^0$ emission at 609 $\mu$m, which peaks at a slightly larger distance from the star and is broader than the CO lines.
This is expected since the C$^0$ emission at 609 $\mu$m has a low energy level and critical density (see Sect. \ref{sect_line_detection_spatial_SED}).
We do not include the C$^0$ emission at 370 $\mu$m since this line is blended with the CO J=7-6 line and it is therefore difficult to extract a map because of the low spectral resolution.

We compare our data to earlier {\it Spitzer} observations of the ${\rm H_2}$ 0$-$0 S(1) rotational line emission (Fig. \ref{fig:250mic} top, right), another major contributor to the gas cooling. 
The first pure H$_2$ rotational lines (e.g. 0$-$0 S(0) and S(1)) have upper states lying $\sim$510~K and $\sim$1015~K above ground
and their critical densities are relatively low even at low temperatures \citep[$n_{\rm crit} < 10^4$ cm$^{-3}$ for T $\ge 100$~K,][]{le-bourlot-et-al-1999}. 
Thus, H$_2$ is expected to probe a warmer and more diffuse gas than the excited CO rotational lines.
In fact, compared to the observed $^{12}$CO and $^{13}$CO emission, the ${\rm H_2}$ emission is shifted slightly towards the star (by about 10'') and appears as an elongated shape parallel to the $^{12}$CO J=7$-$6 and 6$-$5 line emission region (see Figs. \ref{fig:250mic} and \ref{fig:position} bottom right). 
A previous ground-based map of the $^{12}$CO (6$-$5) integrated intensity overlaid on the near-infrared image of the H$_2$ 1$-$0 S(1) emission also shows this shift between CO and H$_2$ \citep[see Fig. 4 in][]{gerin-et-al-1998}. 
We compare our data to the H$_2$ 1$-$0 S(1). 
The H$_2$ 1$-$0 S(1) vibrational emission comes from narrow filaments \citep[width: $\le$5'' or 0.01 pc,][]{lemaire-et-al-1996}, which coincide spatially with filaments seen in HCO$^+$ 1$-$0 (width of 6'') with different densities and velocities \citep{fuente-et-al-1996}.
There was a controversy about the nature of the IR filaments as to whether they are just the border of the cloud or high density narrow filaments as discussed by \citet{fuente-et-al-1996}. 
Comparison between single dish and interferometric HCO$^+$ data suggests that the IR filaments cannot be interpreted as the border of clouds. 
The dense gas is confined into filaments immersed in an atomic or diffuse molecular medium.
\citet{fuente-et-al-1996} suggest that these filaments located along the walls of the cavity could result from the interaction of the bipolar stellar outflow with the cloud.
Figure \ref{fig:250mic} shows that the peak position of the high-J CO emission named position A is slightly shifted to the IR H$_2$ densest filament \citep[$n_{\rm H}$$\approx10^6$~${\rm cm^{-3}}$,][]{fuente-et-al-1996,lemaire-et-al-1996}.
Emission of the intermediate-J CO lines follows the less dense filaments \citep[$n_{\rm H}$$\approx10^5$~${\rm cm^{-3}}$,][]{fuente-et-al-1996,lemaire-et-al-1996}).
The size of the filaments seen in H$_2$ and HCO$^+$ is not resolved with the SPIRE beam.
A beam dilution factor must be taken into account when analysing the excited CO line emission (see discussion in Sect. \ref{SED-7023NW}).

We also compare our data to the $^{13}$CO J=2$-$1 observations with the K\"olner Observatorium f\"ur SubMillimeter Astronomie (KOSMA) 3-meter telescope by \citet{yuan-et-al-2013}. The spatial distribution of this low-excited line differs from the spatial distributions we find for higher excited $^{13}$CO and $^{12}$CO lines. The $^{13}$CO J=2$-$1 lines peak further north, where gas temperatures are lower and where we find a high column density (see Sec. \ref{sec:modBB}).

\subsection{NGC 7023 E}

\begin{center}
\begin{figure*}[th]
\includegraphics[width=0.9\textwidth]{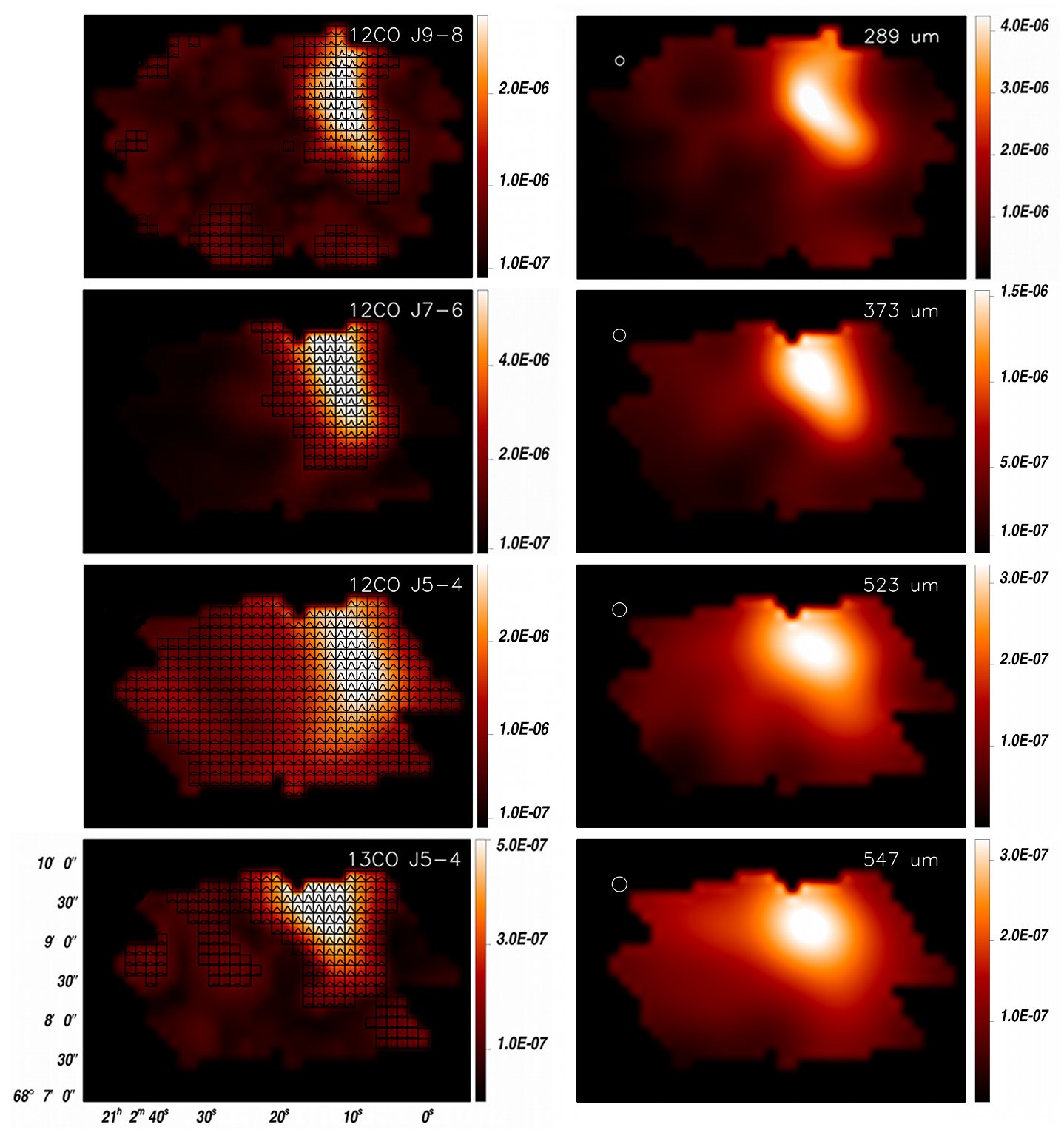}
\caption[]{Left column: Spatial distribution of ${\rm ^{12}CO}$ (J=9$-$8), ${\rm ^{12}CO}$ (J=7$-$6), ${\rm ^{12}CO}$ (J=5$-$4), ${\rm ^{13}CO}$ (J=5$-$4) lines for NGC 7023 E. The bars indicate intensities in ${\rm erg~s^{-1} cm^{-2} sr^{-1}}$. 
Right column: Spatial dust distribution at the wavelength close to the line emission.
The bars give ${\rm \nu I_{\nu}}$ in ${\rm W m^{-2} sr^{-1}}$. All images are presented with the same scale, field, and centre. We include the FWHM of the beam in the dust emission maps: 13.0'', 18.6'', 21.3'', 22.7'' from top to bottom.}
\label{fig:linedistre}
\end{figure*}
\end{center}

In Fig. \ref{fig:linedistre} the emission maps for different ${\rm ^{12}CO}$ and ${\rm ^{13}CO}$ lines and for the dust emission at 289, 373, 523, and 547 $\mu$m (top to bottom) are shown for NGC 7023 E.
In Fig. \ref{fig:250mic} we present the FTS dust map at 250 $\mu$m and the emission of some gas lines included as contours.

The maps show a good spatial correlation between ${\rm ^{12}CO}$ J=9$-$8 emission and dust emission at 250 $\mu$m in a small central region (50'' $\times$ 100'' or 0.1 $\times$ 0.2 pc in size). 
The ${\rm ^{12}CO}$ J=5$-$4 line is observed over the entire PDR. 
The emission of ${\rm ^{13}CO}$ is spatially correlated with the dust emission at wavelengths around 500 $\mu$m in a region shifted towards the north.
This region is close to a core, located a bit further north than the edge of the FTS map, which is seen in ${\rm ^{13}CO}$ J=2$-$1 emission \citep{yuan-et-al-2013}. 
In this region, we find a high column density (see Sec. \ref{sec:modBB}).

In Fig. \ref{fig:position} we plot the gas lines and dust continuum emission profiles along a cut (black line in Fig. \ref{fig:250mic}).
The dust emission at 250 $\mu$m, 500 $\mu$m, and 1.2 mm peaks at the same position at about 220'' from the star.
The emission profiles are relatively broad with a width of about 45$-$55'' ($\sim$0.1 pc), which is about 2$-$3 times the width of the dust emission in NGC 7023 NW.
We also include the distribution of ${\rm H_2}$ rotational emission \citep{habart-et-al-2011} that is shifted by 10$-$15'' towards the star compared to ${\rm ^{12}CO}$ and ${\rm ^{13}CO}$, and by 30'' compared to the dust emission. The dust emission is slightly shifted by 10-15'' to the CO emission but this shift depends on the choice of the cut (see Fig. \ref{fig:250mic}).
The C$^0$ line at 609 $\mu$m peaks at larger distances of 220'' and is broader than the CO lines.

%%%%%%%%%%%%%%%%%%%%%%%%%%%%%%%%%%%%%%%%%%%%%%%%%%%%%%%%%%%%%%%%%%%%%%%%%

\section{Physical properties derived from gas line analysis}\label{SED-7023NW}

In this section, we present the spectral energy distribution of the observed CO lines at specific positions in NGC 7023 NW and NGC 7023 E.
For NGC 7023 NW, we choose three positions of interest marked as A, B, and C in Fig. \ref{fig:250mic}. 
Position A ($\alpha_{\rm 2000}$= 21h 1m 28.3s, $\delta_{\rm 2000}$= 68$^{\circ}$ 10' 22'') is the region where the ${\rm ^{12}CO}$, ${\rm ^{13}CO}$ and dust emission peak, position B ($\alpha_{\rm 2000}$= 21h 1m 32.8s, $\delta_{\rm 2000}$= 68$^{\circ}$ 11' 20.2'') is the region of the northern dust peak and position C ($\alpha_{\rm 2000}$= 21h 1m 36s, $\delta_{\rm 2000}$= 68$^{\circ}$ 10' 55'') is in the elongated bar of intermediate-J ${\rm ^{12}CO}$.
For NGC 7023 E, two positions of interest are chosen and marked as A and B in Fig. \ref{fig:250mic} (bottom).
Position A ($\alpha_{\rm 2000}$= 21h 2m 16s, $\delta_{\rm 2000}$= 68$^{\circ}$ 9' 25'') is the region where the ${\rm ^{13}CO}$ and dust emission peak and position B ($\alpha_{\rm 2000}$= 21h 2m 11s, $\delta_{\rm 2000}$= 68$^{\circ}$ 8' 50'') is in the elongated bar of intermediate-J ${\rm ^{12}CO}$ and 250 $\mu$m dust emission. 

We calculate the integrated line intensities of the ${\rm ^{12}CO}$ and ${\rm ^{13}CO}$ ladders of the observed FTS spectra as described in Sect. \ref{sec:linefit}.
We use the LR data (consistent beam size) to ensure that we compare line intensities from the same area.
The convolution can smooth the small areas of high-J ${\rm ^{12}CO}$ line emission as seen in the previous section and can decrease their intensities, however, these variations are included in the error bars we consider.

\begin{center}
\begin{figure*}[t]
\includegraphics[width=0.75\textwidth]{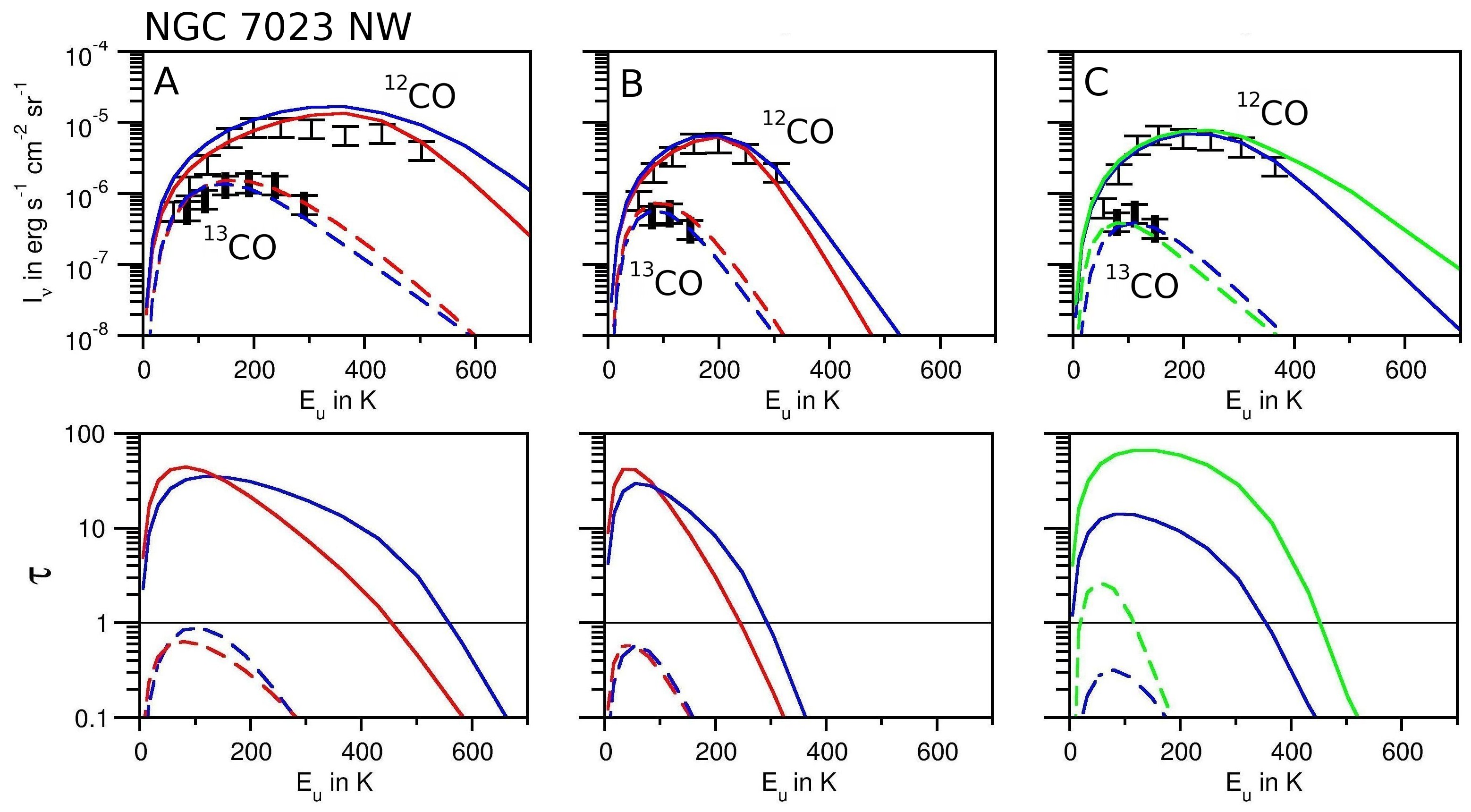}
\caption[]{Top: Integrated line intensities as a function of $E_{\rm u}$ for the positions A, B, and C. The lines indicate the best fits of the observed integrated line intensities with RADEX: red for $n_{\rm H}=10^6$~cm$^{-3}$, blue for $n_{\rm H}=10^5$~cm$^{-3}$, and green for $n_{\rm H}=10^4$~cm$^{-3}$.
Bottom: The optical depth $\tau$ calculated with RADEX versus $E_{\rm u}$ for the positions A, B, and C.
${\rm ^{12}CO}$: thin symbols, solid line, ${\rm ^{13}CO}$: thick symbols, dashed line. The results are summarised in Tab. \ref{tab:lines}.}
\label{fig:specilinw}
\end{figure*}
\end{center}

We analyse the integrated line intensities using the NLTE and (local) radiative transfer code RADEX\footnote{http://www.sron.rug.nl/$\sim$vdtak/radex/index.shtml} \citep{van-der-tak-et-al-2007}.
By solving the statistical equilibrium equations, RADEX applies a simple method of the mean escape probabilities to the (1D) radiative transfer. 
By assuming a uniform density and temperature layer, this simple model can be used to assess the physical conditions in which the observed lines arise: gas density, kinetic temperature, line optical depth, column density, and abundance of the species.
Furthermore, one can check the role of collisions versus the radiative coupling of gas excitation to the intense dust continuum radiation within the PDR.

Assuming a uniform density and temperature layer is simplistic because different phases of gas are mixed along the line of sight and inside the beam area.
The gas temperature could vary very rapidly through the PDR layer, and with our spatial resolution the temperature structure is not spatially resolved.
Moreover, there could be reabsorption by gas along the line of sight, but it is rather unlikely for the observed high-excited lines.
Nevertheless, here we focus our analysis on the intermediate-J and high-J ${\rm ^{12}CO}$ and ${\rm ^{13}CO}$ lines that cannot be reabsorbed by a colder gas, since their energy levels and critical densities are relatively high. 
The high-J CO lines may arise from a warmer layer with a smaller column density than the intermediate-J CO lines. 
Fitting all the lines with only one component is therefore an approximation but gives the average physical conditions of the emission zones in the beam.

We consider a grid of input parameters for the total number of hydrogen nuclei\footnote{The input parameter for RADEX is $n_{\rm H_2}$ and we consider $n_{\rm H}=2\times n_{\rm H_2}$. We present $n_{\rm H}$ for a better comparison with following results.}, $n_{\rm H}$, kinetic temperature, $T_{\rm g}$, CO column density, $N_{\rm CO}$, and beam filling factor, $\eta$.
Gas densities vary in a range between 10$^4$ and 10$^6$ cm$^{-3}$,
kinetic temperatures between 10 K and 500 K, CO column densities between 10$^{15}$ to 10$^{19}$ cm$^{-2}$, and the beam filling factor between 0.01 and 1.
We use the new set of collisional rate coefficients calculated by \citet{yang-et-al-2010} including energy levels up to J = 40 for temperatures ranging from 2~K to 3000~K. 
We assume a standard carbon isotopic ratio $^{12}$C/$^{13}$C of 70 \citep{wilson-1999}.
We use the cosmic microwave background radiation temperature of 2.73 K.
We test the non-LTE model, including the observed intense far-infrared and submm radiation by dust (see Sect. \ref{dust}) as the background source, and find that the effect of the dust continuum emission on the CO lines is negligible. 

We use the observed data for the fitting procedure as follows: 
We fit the slope of the cooling curve of the CO lines with different combinations of kinetic temperature and gas density, which reflect a degeneracy between these two parameters.
For a given combination of kinetic temperature and gas density, we obtain the CO column density, $N_{\rm CO}$, by fitting the $^{12}$CO to $^{13}$CO line ratio that is sensitive to the optical depth, which at the line centre depends on the ratio of the column density to the line width. 
For reasons of simplicity, we assume a constant width for all the lines measured in the FTS cubes.
We assume $\Delta \rm{v} = 1.5~{\rm km~s^{-1}}$ in agreement with HIFI observations towards NGC 7023 NW (\citet{joblin-et-al-2010} and Bern\'e et al., in prep.) showing a turbulent line width of about $\Delta \rm{v} = 1-2~{\rm km~s^{-1}}$ for the observed profiles of the high $^{12}$CO and $^{13}$CO lines. 
We assume that the beam filling factor is constant for all of the lines.
For each gas density, our fitting procedure yields the best fit corresponding to the minimum $\chi^2$.

In order to determine the most likely local gas density for the different positions, we calculate the length along the line of sight, $l$, of the emission layer with $l\sim 2\times N_{\rm H_2}/n_{\rm H}$, where $N_{\rm H_2}$ is the beam averaged column density, and compare it to the PDR projected size. 
We assume that the length along the line of sight should not be much smaller than the projected emission width and not much larger than the projected emission extent, where the projected width and extent are the dimensions of the PDRs in the plane of the sky as derived in Sec. \ref{spatial-7023NW}.
In order to estimate $l$, we convert the CO into H$_2$ column densities taking a relative $^{12}$CO abundance to H$_2$ of 10$^{-4}$,
however, this ratio varies widely in the PDR from very low values up to $2.6 \times 10^{-4}$.
Variation in the CO to H$_2$ abundance ratio can result from photodissociation and from CO evaporation of ice mantles of dust grains, which are hotter than about 20$-$25 K.
At the peak position of the intermediate-J CO emission, PDR models \citep{le-petit-et-al-2006} with $G_0=50-10^3$ and $n_{\rm H}=10^4-10^6$~cm$^{-3}$ predict a value of about 10$^{-4}$.
This assumption will be discussed in more detail in Section \ref{density} where we derive the CO to H$_2$ ratio using the total hydrogen column density obtained from the dust.

\begin{table*}[t]
\caption{The results with RADEX for the considered positions A, B, C for NGC 7023 NW (see Fig. \ref{fig:specilinw}) and for the positions A and B for NGC 7023 E (see Fig. \ref{fig:speciline}). Results that give a reasonable physical length along the line of sight are shown in bold face.}
\begin{tabular}{l|ccccc|ccccc|ccccc}
\hline
        $n_{\rm H}$         \T & T        & ${N_{\rm CO}}$ & l       &  $\eta$   & $\chi^2$  &   T     & ${N_{\rm CO}}$  & l           &  $\eta$   & $\chi^2$   &     T        & ${N_{\rm CO}}$ & l          & $\eta$ & $\chi^2$ \\
     $[{\rm cm^{-3}}]$ \B & [K]     & [cm$^{-2}$]               & [pc]     &        &  &  [K]  & [cm$^{-2}$]                & [pc]         &         & &      [K]    & [cm$^{-2}$]               & [pc]       &  &  \\           
\hline
NW \T &   \multicolumn{5}{c|}{A} &  \multicolumn{5}{c|}{B} &  \multicolumn{5}{c}{C} \\
\hline
$10^4$  \T& 160   & $1 \times 10^{19}$   &   6.5   &  0.1 & 26.6& 70   & $5 \times 10^{18}$    &  3.2    &  0.2 & 10.7 & {\bf 120} & ${\bf 4 \times 10^{18}}$    & {\bf 2.6}  & {\bf 0.1} & {\bf 5.6} \\
${5 \times 10^4}$  &{\bf 130}   & $ {\bf 3 \times 10^{18}}$   &  {\bf 0.4}    &  {\bf 0.1} & {\bf 16.9}&    &     &     &   &  &    &  &  \\
${10^5}$  & {\bf 110}  & ${\bf 2.5 \times 10^{18}}$  & {\bf 0.2}  & {\bf 0.1} & {\bf 12.1} & {\bf 50}  & ${\bf 1 \times 10^{18}}$     & {\bf 0.07}      & {\bf 0.2}   & {\bf 2.5} & {\bf 80}  & ${\bf 7 \times 10^{17}}$    & {\bf 0.05}      & {\bf 0.1} & {\bf 2.1} \\
${5 \times 10^5}$  & {\bf 70}   & ${\bf 2 \times 10^{18}}$   & {\bf 0.03}    &  {\bf 0.1} & {\bf 1.6} &  &     &     &   &  &    &  &  \\
${10^6}$  \B& {\bf 65}    & ${\bf 2 \times 10^{18}}$   & {\bf 0.01}  & {\bf 0.1} & {\bf 1.3} &   {\bf 33}    & ${\bf 1 \times 10^{18}}$   & {\bf 0.007}   & {\bf 0.3} & {\bf 1.9} &   50    & $5 \times 10^{17}$   &   0.003  & 0.2 & 3.0 \\
\hline
\hline
E \T &   \multicolumn{5}{c|}{A} &  \multicolumn{5}{c|}{B} &  \\
\hline
$10^4$  \T& {\bf 55}   & ${\bf 4 \times 10^{18}}$    & {\bf 2.6}  &  {\bf 0.2}  & {\bf 2.1} & {\bf 50}  & ${\bf 4 \times 10^{18}}$     & {\bf 2.6}  &  {\bf 0.1}  & {\bf 4.1} & & & & \\
${5 \times 10^4}$  & {\bf 50}   & ${\bf 1 \times 10^{18}}$   & {\bf 0.1}    & {\bf 0.2} & {\bf 1.0} &  &     &     &   &  &    &  &  \\
$10^5$  & {\bf 47}  & ${\bf 5 \times 10^{17}}$   & {\bf 0.03}       &  {\bf 0.3} & {\bf 3.7}  & {\bf 47}   & ${\bf 5 \times 10^{17}}$  & {\bf 0.03}       & {\bf 0.2} & {\bf 0.3} & & & & \\
$5 \times 10^5$  & 35   & $5 \times 10^{17}$   &  0.01    &  0.3 & 0.5 &  &     &     &   &  &    &  &  \\
$10^6$  \B& 32   & $5 \times 10^{17}$  &  0.003  & 0.4 & 2.1 & 32  & $4 \times 10^{17}$  &  0.003   & 0.3 & 1.6 & & & & \\
\hline
\end{tabular}
\label{tab:lines}
\end{table*}

\begin{center}
\begin{figure}[t]
\includegraphics[width=0.5\textwidth]{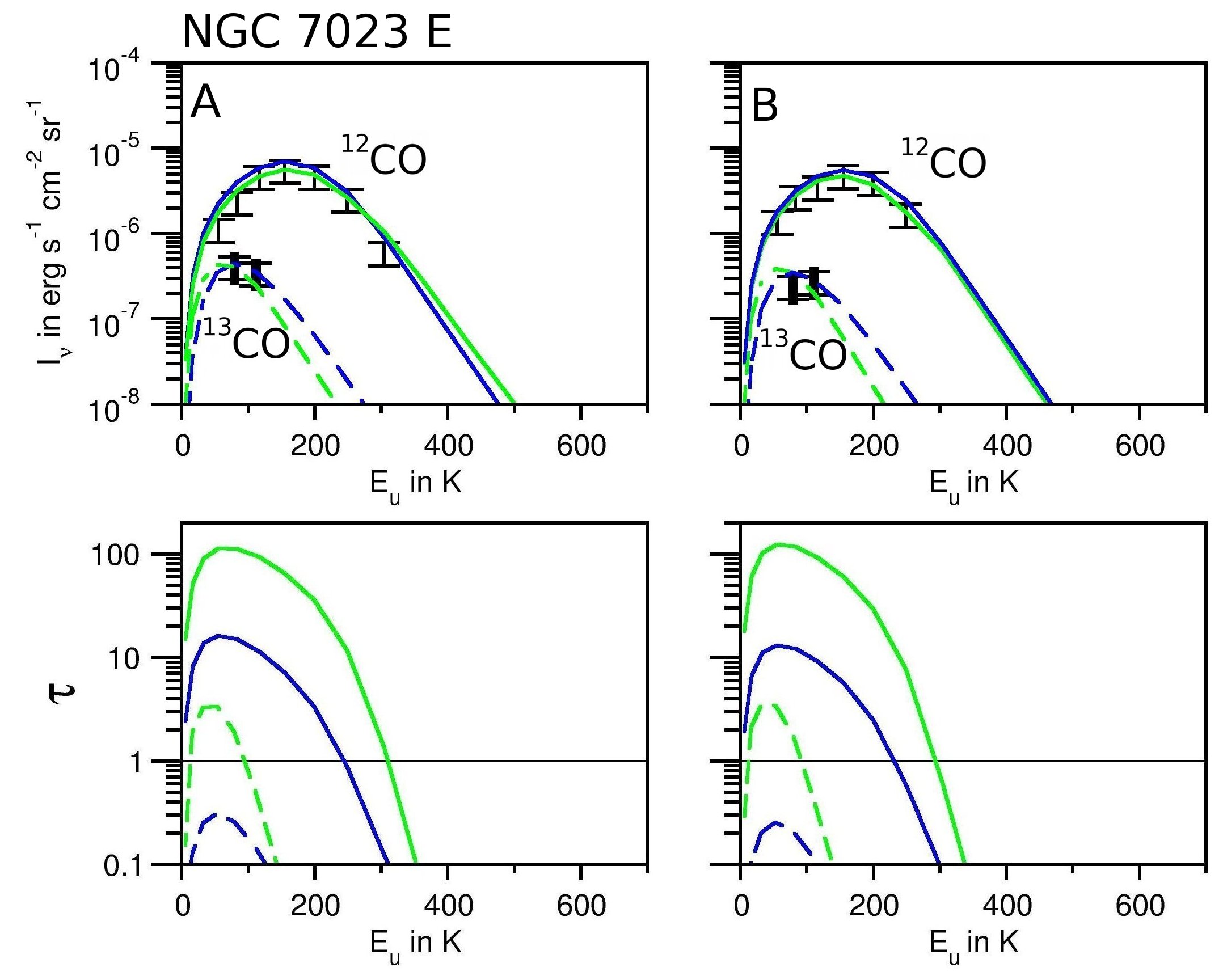}
\caption[]{
Same as Fig. \ref{fig:specilinw} for NGC 7023 E for the positions A and B. The results are summarised in Tab. \ref{tab:lines}.}
\label{fig:speciline}
\end{figure}
\end{center}

\subsection{NGC 7023 NW}

Figure \ref{fig:specilinw} (top row) shows the RADEX results with the observed integrated line intensities of $^{12}$CO and $^{13}$CO in the three positions A, B, and C.
The line intensities can be {fitted} with several combinations of $n_{\rm H}$, $T_{\rm g}$, $N_{\rm CO}$, and $\eta$, since small $n_{\rm H}$ and large $T_{\rm g}$ lead to similar results as high $n_{\rm H}$ and small $T_{\rm g}$.
We therefore present the results for values between $n_{\rm H}=10^4$~cm$^{-3}$ and $n_{\rm H}=10^6$~cm$^{-3}$, which are summarised in Tab. \ref{tab:lines}. 

For position A, we exclude the case of $n_{\rm H}=10^4$~cm$^{-3}$, where we obtain a large CO column density that leads to a length along the line of sight of 6.4 pc, compared to a width of about $\lesssim$0.04$-$0.05~pc and a projected emission extent of about $\sim$0.1~pc (see Sect. \ref{spatial-7023NW}).
This is physically unlikely since it corresponds to about 60 times the projected extent of the high-J CO emission in the plane of the sky and at least about 150 times the projected width.
Reasonable lengths are obtained for a gas density between $n_{\rm H}=5 \times 10^4~{\rm cm^{-3}}$ and $n_{\rm H}=10^6~{\rm cm^{-3}}$.
For these densities, we obtain a beam filling factor of $\eta$ = 0.1 which leads, in the case of a simple elongated\footnote{The beam filling factor $\eta$ can be written as the ratio between the line emission area to the beam area. In a very simple geometrical view of an elongated emission, the line emission area can be estimated by the product between its width and the beam FWHM. The width can thus be written as $\eta \times (\Omega) / ({\rm FWHM})$ with $\Omega$ the beam area. For the low resolution cube the beam area and FWHM are 3.2$\times 10^{-8}$~sr and 42'', respectively.} emission, to a width of about 0.01~pc. This is in agreement with the observed projected width.
Finally, we exclude higher gas densities since they yield unrealistic small lengths. 
We find temperatures between 65 and 130 K, depending on the gas density.  
In all cases the CO column density is relatively high at $2-3 \times 10^{18}$ ${\rm cm^{-2}}$.
This high CO column density is needed to fit the partly small ${\rm ^{12}CO}$/${\rm ^{13}CO}$ ratios indicating optical thickness.
We find a ${\rm ^{12}CO}$/${\rm ^{13}CO}$ J=6$-$5 ratio of about 2$\pm$1, which is very small compared to the standard isotopic ratio\footnote{The average local ISM carbon isotope ratio is about 70 \citep{wilson-1999}.} and indicates that ${\rm ^{12}CO}$ J=6$-$5 is highly optically thick.
The line ratio ${\rm ^{12}CO}$/${\rm ^{13}CO}$ increases with increasing $E_{\rm u}$ and is about 10$\pm$3 for the J=10$-$9 line showing that the optical thickness decreases with J, however, the ratio that indicates optical thickness is still small.
The optical depths calculated by RADEX as a function of $E_{\rm u}$ are presented in Fig. \ref{fig:specilinw} (bottom) for $^{12}$CO and $^{13}$CO. 

The H$_2$ rotational lines provide rotational excitation temperatures of $\sim$300 to 1000 K \citep{fuente-et-al-1999,fleming-et-al-2010} and a column density {of 0.5$-$5 $\times 10^{20}~{\rm cm^{-2}}$.
This confirms that H$_2$ probes a warmer region closer to the star.
Considering ground-based observations of the ${\rm ^{13}CO}$ J=3$-$2 line by \cite{gerin-et-al-1998} and the ${\rm ^{13}CO}$ J=5$-$4 lines from FTS observations, we find gas temperatures of around 20 K and a ${\rm ^{12}CO}$ column density of around $N=3.5 \times 10^{17}$ ${\rm cm}^{-2}$ assuming a standard isotopic ratio ${\rm ^{12}C}$/${\rm ^{13}C}$ of 70.
It seems that the ${\rm ^{13}CO}$ J=3$-$2 traces cold gas in dense and shielded regions that are not influenced by stellar radiation (see Sec. \ref{sec:east}).

For position B, we exclude a gas density of $n_{\rm H}=10^4~{\rm cm^{-3}}$ since the length along the line of sight is around 30 times larger than the projected length.
For this position, where we detect no high-J ${\rm ^{12}CO}$ lines and only three ${\rm ^{13}CO}$ lines, 
we derive temperatures of $T=33-50$ K that are significantly lower than for the position A. 
The derived CO column density is a factor 2$-$3 smaller than for position A and the ratio between the ${\rm ^{12}CO}$ and ${\rm ^{13}CO}$ J=6$-$5 line is slightly larger by about 3$\pm$1.8.

For position C, we consider gas densities between $n_{\rm H}=10^4~{\rm cm^{-2}}$ and $n_{\rm H}=10^5~{\rm cm^{-2}}$, while $n_{\rm H}=10^6~{\rm cm^{-2}}$ leads to a very small length of the PDR (10 times smaller than the projected width).
This latter case would only be possible if the CO to H$_2$ abundance is about $10^{-5}$.
The derived temperature is between 120 and 80 K. 
The line ratio ${\rm ^{12}CO}$/${\rm ^{13}CO}$ J=6$-$5 is about five times larger than in positions A and B.

\subsection{NGC 7023 E}\label{sec:east}

For the two positions A and B in NGC 7023 E, the results for the RADEX calculations are also presented in Tab. \ref{tab:lines}. The fits to the observed CO lines are presented in Fig. \ref{fig:speciline} (top row).

For both positions, good fits are obtained for an intermediate gas density between $n_{\rm H}=10^4~{\rm cm^{-3}}$ and $n_{\rm H}=10^5~{\rm cm^{-3}}$ for which the length along the line of sight is comparable to the projected width and extent of around $\le$0.1 to 0.2 pc, respectively. 
For $n_{\rm H}=10^6~{\rm cm^{-3}}$ we obtain a rather small length along the line of sight (around 30 times smaller than the projected width) so that we exclude this case. 
We find colder temperatures of around 50 K compared to NGC 7023 NW, which is due to the fact that NGC 7023 E is further away from the central star and the incident UV flux is more diluted and the gas is less excited.
The gas temperature at the two positions do not show such a large variety as seen in NGC 7023 NW,
which indicates that the radiation field is less rapidly attenuated in this lower density and more extended region. 
The latter is further indicated by larger filling factors compared to NGC 7032 NW.
For both positions, the $^{\rm 12}$CO lines are optically thin for $E_{\rm u}>$250 K (Fig. \ref{fig:speciline} bottom). 
The column density is similar to that in NGC 7023 NW, where high-J and intermediate-J lines are optically thick.

The analysis of the H$_2$ rotational lines yields rotational excitation temperatures of $\sim$270$-$330 K \citep{habart-et-al-2011,fleming-et-al-2010} and a column density between $3 \times 10^{20}~{\rm cm^{-2}}$ and $1.5 \times 10^{21}~{\rm cm^{-2}}$. 
This confirms that H$_2$ probes a warmer region closer to the star.

From ground-based observations of the ${\rm ^{13}CO}$ J=2$-$1 line by \cite{yuan-et-al-2013} together with the ${\rm ^{13}CO}$ J=5$-$4 lines from FTS observations, we find gas temperatures of around 20 K and a ${\rm ^{12}CO}$ column density of around $N=2.8 \times 10^{17}$ ${\rm cm}^{-2}$ assuming a standard isotopic ratio ${\rm ^{12}C}$/${\rm ^{13}C}$ of 70.
Since the derived temperature and column density are similar for NGC 7023 NW and NGC 7023 E, the ${\rm ^{13}CO}$ J=3$-$2 line seems to be rather independent of the radiation field. 
The ${\rm ^{13}CO}$ J=3$-$2 lines trace cold gas in shielded regions, which are distinct from the regions traced with the higher-excited CO lines.

\begin{center}
\begin{figure*}[t]
\includegraphics[width=0.99\textwidth]{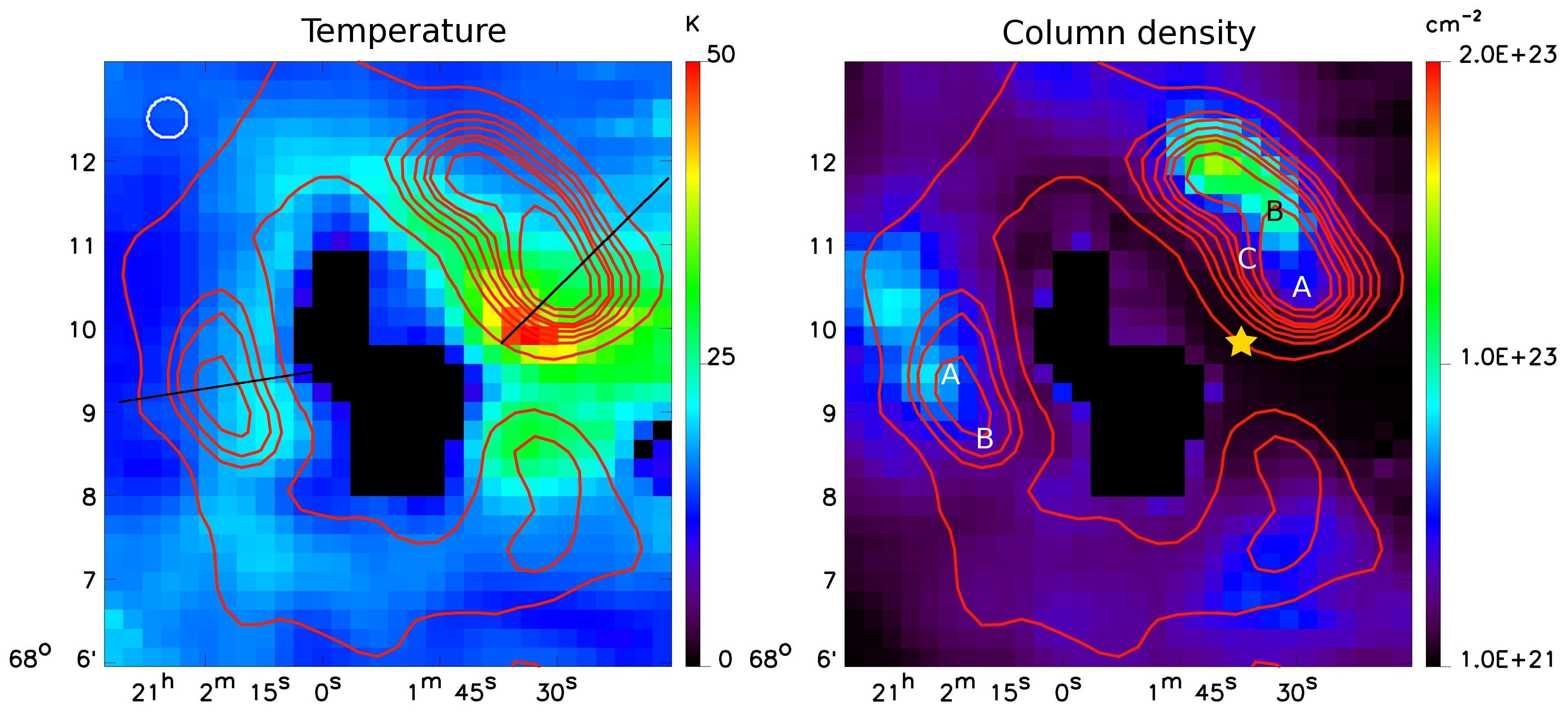}
\caption[]{The dust temperature (left) and column density, $N{_{\rm H}}$, (right) for NGC 7023 for a fixed $\beta$=2. The red contours show the 250\,$\mu$m emission at 300, 500, 600, 750, 900, 1000, 1250, and 1500 MJy sr$^{-1}$. Maps are convolved to the 500 $\mu$m beam with a FWHM of 35.1'' (white circle).}
\label{fig:heddy}
\end{figure*}
\end{center}

%%%%%%%%%%%%%%%%%%%%%%%%%%%%%%%%%%%%%%%%%%%%%%%%%%%%%%%%%%%%%%%%%%%%

\section{Physical properties derived from dust continuum analysis}\label{dust}

In this section, we focus on the analysis of dust emission as an independent tracer.
Observational data obtained with the Infrared Array Camera (IRAC), Multiband Imaging Photometer (MIPS) ({\it Spitzer}), PACS, and SPIRE photometers (SUPREME method is not used but standard data processing with naive gridding method) and IRAM, convolved to the 500 $\mu$m beam with a FWHM of 35.1'', are used to obtain the physical properties for NGC 7023 NW and E. 
These data offer a large wavelength range from 3.6 \,$\mu$m to 1.2 mm, covering the entire infrared emission of grains, and are described in \citet{abergel-et-al-2010}. 
Physical properties within the PDRs are determined using two complementary approaches: modified black body fits and radiative transfer calculations.

\subsection{Modified black body fits}      
\label{sec:modBB}

As a first approach, we focus our study on the big grains (BGs), whose emission is observed at wavelengths longer than $\sim$100 $\mu$m.
These BGs are in thermal equilibrium so that we can use a single modified black body emission law to fit the observations and determine the dust temperature, $T_{\rm{d}}$, and the hydrogen column density, $N_{\rm{H}}^{\rm{BB}}$:
\begin{equation}
I_\lambda=\tau_{\lambda_{0}}\times\left(\frac{\lambda}{\lambda_0}\right)^{-\beta}\times B_\lambda(T_{\rm d}),
\label{eq:BB_fit}
\end{equation}
where $\tau_{\lambda_{0}}$ is the dust optical depth at wavelength $\lambda_0$, $\beta$ is the spectral emissivity index and $B_\lambda$ is the Planck function. 
The fits are performed for each pixel of the data cube and we consider only wavelengths corresponding to the emission of BGs, i.e. PACS 160\,$\mu$m and SPIRE 250, 350, and 500\,$\mu$m. 
The errors considered in this approach are statistical noise (signal-to-noise ratio larger than 40 in the worst case at 500\,$\mu$m).
The free parameters are $\tau_{\lambda_{0}}$ and $T_{\rm{d}}$. The spectral emissivity index $\beta$ is fixed to 2 since we are not interested in $\beta$-T anti-correlation effects. With $\beta$ as a varying parameter, we find a slightly larger temperature range.

These modified black body fits allow us to map the temperature and the column density variations. The latter is computed from the optical depth at 250 $\mu$m according to:
\begin{equation}
N_{\rm{H}}^{\rm{BB}}=\frac{\tau_{250\,{\rm \mu m}}}{\kappa_{250\,{\rm \mu m}} \,\mu \,m_{\rm{H}}},
\label{eq:nhbb}
\end{equation} 
where $\kappa_{250~{\rm \mu m}}=0.051$ cm$^2$g$^{-1}$ is the dust emissivity for the diffuse ISM derived by \citet{compiegne-et-al-2011}, $\mu=1.33$ is the mean atomic mass, and $m_{\rm{H}}$ is the hydrogen mass. This simple calculation assuming a constant dust emissivity and neglecting the radiative transfer effects provides an initial estimate of the temperature and column density. 

The temperature and column density maps are shown in Fig. \ref{fig:heddy} \citep[for details see][]{arab-et-al-2012b} and the values for the chosen positions in both PDRs are summarised in Tab. \ref{table:fits}. 
The error on $N_{\rm{H}}^{\rm{BB}}$ given here is derived from the fits, however, assumptions we made in this approach could bias the column density. 
In NGC 7023 NW, we find dust temperatures of about 50$-$55\,K right in front of the edge of the cloud.
Towards position C, representing another part of the edge of the cloud (see the visible map in Fig. \ref{fig:full}), the dust temperature decreases, which could be due to a geometrical effect, i.e. position C is further away from the star (see the visible map in Fig. \ref{fig:full}).
The dust temperature decreases to around 30$-$35\,K when moving away from the star and entering the PDR and decreases further to around 20 K close to the molecular cloud. 
The column density at the position of largest temperature in NGC 7023 NW is rather small and increases when entering the PDR.
The maximal H column density of $1.5\times10^{23}$\,cm$^{-2}$ is reached at around 2' northeast from the temperature peak. 
This position is not observed with SPIRE FTS.
In NGC 7023 E, we obtain dust temperatures of around 20\,K, which slightly decrease with distance from the star. The column density $N_{\rm H}$ increases slightly with distance from the star to up to $8\times10^{22}$\,cm$^{-2}$. 
The regions of high column density coincide spatially with cores observed in $^{13}$CO J=2$-$1 emission \citep{yuan-et-al-2013}.

\begin{table}
\caption{Dust temperature $T_{\rm d}$ and column density $N_{\rm{H}}^{\rm{BB}}$ at the positions A, B, and C in NGC 7023 NW and at the positions A and B in NGC 7023 E. }             
\label{table:fits}               
\begin{tabular}{l|ccc}  
\hline
NW \T\B & A & B & C\\
\hline       
$N_{\rm{H}}^{\rm{BB}}$ [cm$^{-2}$] \T\B   &	 (4$\pm$0.4)$\times 10^{22}$	& (9.8$\pm$1.0)$\times 10^{22}$ 	& (2.8$\pm$0.3)$\times 10^{22}$ 	 \\ 
$T_{\rm{d}}$ [K] \T\B	  & 30.0$\pm$3.4 	& 20.1$\pm$1.2 & 30.3$\pm$2.6  \\  
\hline           
\hline       
E  \T\B & A & B & \\
  \hline
$N_{\rm{H}}^{\rm{BB}}$ [cm$^{-2}$] \T\B   & (4$\pm$0.5)$\times 10^{22}$	& (7$\pm$0.9)$\times 10^{22}$  & \\ 
$T_{\rm{d}}$ [K] \T\B	& 20.5$\pm$0.9 & 17.2$\pm$0.6  & \\  
\hline
\end{tabular}
\end{table}

\subsection{Radiative transfer modelling}

As a second approach, we use a 1D radiative transfer model combined with DustEM as described in \citet{arab-et-al-2012} based on the dust model of \citet{compiegne-et-al-2011} assuming an edge-on geometry. 
The dust model of \citet{compiegne-et-al-2011} describes the observed spectral energy distribution and extinction of the diffuse ISM with PAH (65 ppm of C), very small carbonaceous grains (14 ppm of C), big carbonaceous grains (121 ppm of C), and big silicate grains (45 ppm of Si).
Each PDR is described by a plane parallel slab defined by its density profile, which is found by fitting the modelled emission with the observations. 
We assume a symmetric power-law profile with a density plateau: $n(z)=n_0 (\frac{z}{z_0})^{\alpha}$, where $z$ is the distance to the star, $n_0$ the maximum density, $z_0$ the distance where the maximum density is reached, and $\alpha$ the power-law.
The PDR is completely defined by these parameters together with the width of the density plateau and the length of the PDR along the line of sight, $l_{\rm{PDR}}$. This last parameter is directly linked with the column density $N_{\rm{H}}$. 
For further details on the model, see Arab et al. (in prep.). 

The radiative transfer modelling is carried out along a cut shown in Fig. \ref{fig:250mic} and \ref{fig:heddy} towards the positions A in both PDRs. We present the results for the three maximum densities $n_{0}=10^4$, $n_{0}=10^5$, and $n_{0}=10^6$~${\rm cm^{-3}}$ with a power-law index, $\alpha$, of 1.5, 2.0, and 3.5, respectively, for NGC 7023 NW and with $\alpha=2$ for all densities in NGC 7023 E.
The density profiles for these three cases are presented in Fig. \ref{fig:denprof}.
The density profile of NGC 7023 E is broader than that of NGC 7023 NW, where the latter is marginally resolved and might be even sharper.
The slope towards the star, the width of the plateau, and a decrease towards the molecular cloud are necessary to fit the observations. 
However, the model cannot constrain the slope of the decreasing density, and we therefore show this part of the density profile in dashed lines in Fig. \ref{fig:denprof}. 
We assume a symmetric profile for simplicity, although the gas density does not decrease to zero inside the cloud and instead should be at least $10^4$ ${\rm cm^{-3}}$ (see \citet{fuente-et-al-1990} and Fig. 11).

In Tab. \ref{tab:dustrtc}, we present for each density profile, the density at the CO emission peak, $n^{\rm e}_{\rm H}$, the visual extinction integrated from the edge to the emission peak, ${\rm A_V^e}$, the column density along the line of sight where the CO emission peaks, ${N^e_{\rm H}}$, and the length along the line of sight, $l_{\rm{PDR}}$. 
The results show that with increasing density, the column density increases and the length along the line of sight decreases.

The best fits to the observations (Fig. \ref{fig:dustres}) are obtained for both PDRs with density profiles with a maximum density of $n_0 = 2\times10^5$~${\rm cm^{-3}}$ and a slope $\alpha=2$. 
The width of the density profile is larger for NGC 7023 E than for NGC 7023 NW and the maximum density is reached closer to the edge in NGC 7023 NW (0.03 pc) than in NGC 7023 E (0.14 pc). 
These density profiles are shown as black curves in Fig. \ref{fig:denprof} (top middle versus distance in pc (arcsec) and bottom versus $A_{\rm V}$) and are associated with a length along the line of sight of 0.2 pc for NGC 7023 NW and 0.15 pc for NGC 7023 E.
We find a good agreement comparing the values of $N^{\rm {e}}_{\rm H}$ and $N^{\rm {BB}}_{\rm H}$ for positions A in NGC 7023 NW and NGC 7023 E. 

The fits presented in Fig. \ref{fig:dustres} (blue curves) are for NGC 7023 NW (right) and NGC 7023 E (left).
In order to fit the observations, we have to include variations in the dust properties relative to the original dust model of \citet{compiegne-et-al-2011}, which is based on the diffuse emission at high galactic latitudes (see red curves in Fig. \ref{fig:dustres}).
For NGC 7023 NW, the PAH abundance is decreased by a factor of 2 and the BG emissivity is increased by a factor of 2.
In the case of NGC 7023 E, only the BG emissivity is increased by a factor of 2.
These variations provide a decrease in the 3.6 $\mu$m emission as it is observed, compared to the original dust model of \citet{compiegne-et-al-2011}.
The decrease in the PAH emission is degenerated with the length along the line of sight and the BG emissivity.
The change in the PAH abundance and in the BG emissivity reflects the evolution of dust from the diffuse to dense regions in the ISM, as described in detail in \citet{arab-et-al-2012}. 

However, deviations between observations and our model occur at 70, 350, 500, and 1200 $\mu$m, which are likely to result from evolutionary processes of dust grains not taken into account with the simple dust model considered here.
The decrease in the 70 $\mu$m emission is seen in observational data of denser regions \citep[see e.g.][]{stepnik-et-al-2003}.
Very small grains, which are responsible for the emission at this wavelength in the diffuse ISM, could coagulate onto the surface of BGs, so that the 70 $\mu$m emission decreases.
The deviations at wavelengths longer than 350 $\mu$m are possible due to changes in the spectral index, which is found to increase from the diffuse to dense regions in the ISM \citep[see e.g.][]{sadavoy-et-al-2013}. 
An increase in the spectral index could occur due to coagulation and/or accretion processes.
It is beyond the scope of this paper to investigate the variations of the modelled results with the evolution of dust. 
A study of these evolutionary processes of dust is in preparation (K\"ohler et al., in prep.), including the new dust model from \citet{jones-et-al-2013}. 

However, the discrepancies that are discussed above are marginally linked to the density profile and tests show that the density profiles appear to be robust against (realistic) changes in the dust model. 
The derived density profile at the edge of NGC 7023 NW is also in good agreement with that determined by \citet{pilleri-et-al-2012}.
Their study focusses on PAHs at the edge of the clouds, where we find the same slope of the density profile using the dust model from \citet{compiegne-et-al-2011}, which includes PAHs and the other dust populations.

In summary, with the radiative transfer modelling we derive the density profile that yields a steep gradient at the edge of the cloud and a subsequent decrease inside the cloud. 
The density profiles derived here are comparable to the profiles found for other PDRs, such as the Horsehead nebula and the Orion Bar \citep{arab-et-al-2012b}.
The length along the line of sight varies with the maximum density. 
As for the CO analysis, the length along the line of sight should not be much smaller than the projected emission width and not much larger than the projected emission extent. For NGC 7023 NW, we can therefore exclude the case of $n_0=10^4$ ${\rm cm^{-3}}$. 
For $n_0=10^5$ ${\rm cm^{-3}}$, the length is comparable to the projected emission extent resulting in a flat extended geometry, while for $n_{\rm 0}=10^6$ ${\rm cm^{-3}}$ it is comparable to the projected emission width resulting in a cylindric geometry. 
For NGC 7023 E, we exclude the case of $n_0=10^6$ ${\rm cm^{-3}}$.
For $n_0=10^5$ ${\rm cm^{-3}}$, the obtained length is between the projected width and extent.
We do not exclude the case of $n_0=10^4$ ${\rm cm^{-3}}$ although the length is three times larger than the projected emission extent.

\subsection{Discussion}

In this section, we compare the results derived with the modified black body calculations and the radiative transfer modelling. We compare these results further to interferometric millimetre observations and end with a discussion on the origin of the derived structures. 

With the modified black body and the radiative transfer modelling approaches, we are able to derive the column density profile of NGC 7023 NW and E. 
The $N_{\rm H}$ column density derived with both approaches is shown in Fig. \ref{fig:Nh}. 
The results look rather different at first view, and in particular, the maximum values for $N_{\rm H}$ for the NW PDR show large deviations.
The deviation in maximum value is probably even larger when we take into account that for the column density derived with the modified black body we use the $\kappa_{250~{\rm \mu m}}$ from the diffuse ISM (see Sec.\ref{sec:modBB}), while for the radiative transfer calculations we increase the BG emissivity by a factor of 2.
At second view, however, we find a similar maximum position, a similar width, and a similar slope of increase, which is steeper for NGC 7023 NW compared to NGC 7023 E.

Using millimetre-interferometric data towards NGC 7023 NW, \citet{fuente-et-al-1996} reveal high density filaments with a width of $\sim$0.01 pc observing HCO$^+$.
Our results are in agreement with this study although the obtained width in our study is around five times larger, in part due to the lower resolution of the {\it Herschel} SPIRE observations.
Considering our strong assumptions about the geometry, we derive the main trend of the density profile.

The high densities may result from a bipolar outflow interaction with the cloud \citep{fuente-et-al-1996}, a shock driven by the high pressure FUV-heated layer \citep[e.g.][]{gorti-hollenbach-2002,hosokawa-inutsuka-2006}, or a phenomenon comparable to evaporating molecular clouds in blister HII regions \citep{bedijn-et-al-1981}.
Considering that the gas in the cavity would be mostly neutral and the thermal pressure is not higher than in the PDR (see Sect. \ref{pressure}), the study by \citet{hosokawa-inutsuka-2006} does not appear to be applicable.
It is also likely that prior to star formation in the molecular cloud density inhomogeneities with dense spots (filaments or clumps) were present.
Our study shows that the density is increased by up to $\sim$10 - 100 times the ambient gas density in a layer with a geometrical thickness of about 0.05$-$0.2 pc.
Wide-field molecular cloud studies by \citet{andre-et-al-2011} and \citet{arzoumanian-et-al-2013} show that dense filaments have a characteristic width of around 0.1 pc. 
These filaments could be formed by turbulence. 
The energy dissipation length is on the order of 0.1 pc in regions that are well shielded from UV \citep{hennebelle-et-al-2013}.
The filaments we observe in NGC 7023 could therefore be formed prior to the formation of HD 200775.

\begin{table}[t]
\caption{The results for the radiative transfer calculations for dust in NGC 7023 NW and in NGC 7023 E where $n_{\rm H}^{\rm e}$ is the density at the CO emission peak, $n_0$ is the maximum density, ${\rm A_V^e}$ is the visual extinction perpendicular to the line of sight integrated from the edge to the emission peak, ${N^e_{\rm H}}$ is the column density along the line of sight at the emission peak, $l_{\rm{PDR}}$ is the length along the line of sight, and ${N_{\rm ^{12}CO}/N^e_{\rm H_2}}$ is the $^{12}$CO to H$_2$ abundance ratio at the emission peak. We derive $N_{\rm CO}$ from the RADEX analysis (see Tab. \ref{tab:lines}) and $N_{\rm H_2}$ from ${N^e_{\rm H}}/2$. The results for the best fit are shown in blue. The table entries in bold face indicate results for physically reasonable lengths along the lines of sight.}
\begin{tabular}{lccccc}
\hline
$n_{\rm H}^{\rm e}$  	    		&        $n_0$               \T     &  ${\rm A_V^e}$	  	& ${N^e_{\rm H}}$          & $l_{\rm{PDR}}$  	&   ${N_{\rm ^{12}CO}/N^e_{\rm H_2}}$   \\
     $[{\rm cm^{-3}}]$ \B \T     & $[{\rm cm^{-3}}]$ 		&  					& [cm$^{-2}$]                    	   & [pc]  	&       \\           
\hline
 \multicolumn{5}{l}{NGC 7023 NW}     \T\B    \\
\hline
$10^4$  		   		     & $10^4$  \T                               &   $\sim$0.2 		& $ 2 \times 10^{22}$         	   & 0.6   	           &    -    \\
${\bf 6 \times10^4}$  	     &${\bf 10^5}$  \T                        &   {\bf $\sim$1}	        & $ {\bf 4 \times 10^{22}}$            & {\bf 0.2}          &    ${\bf 1.6\times 10^{-4}}$   \\
${\bf \blau 1.2 \times 10^{5}}$  &${\bf \blau 2 \times 10^5}$  \T  &   {\bf \blau $\sim$2}  & $ {\bf \blau 7 \times 10^{22}}$  & {\bf \blau 0.2}  &    ${\bf \blau 7 \times 10^{-5}}$   \\
${\bf 4\times10^5}$  		     &${\bf 10^6}$  \T\B                     &    {\bf $\sim$4}	        & $ {\bf 4 \times 10^{22}}$           & {\bf 0.03}         &    ${\bf 1\times 10^{-4}}$\\
\hline
 \multicolumn{5}{l}{NGC 7023 E}    \T\B     \\
\hline
${\bf 5 \times 10^3}$       	& ${\bf 10^4}$  \T                          &   {\bf $ \sim$0.2} 		& $ {\bf 9 \times 10^{21}}$        	   & {\bf 0.6}   	      &     - 	\\
${\bf 5 \times 10^4}$       	&${\bf 10^5}$  \T    			    &   {\bf $\sim$2}		& $ {\bf 6  \times 10^{22}}$           & {\bf 0.4}     	      &    ${\bf 3 \times 10^{-5}}$  \\
${\bf \blau 5 \times 10^4}$  &${\bf \blau 2 \times 10^5}$  \T    &   {\bf \blau $\sim$2}	& $ {\bf \blau 2 \times 10^{22}}$  & {\bf \blau 0.15} &    ${\bf \blau  1 \times 10^{-4}}$     \\
$5\times 10^5$      	 	&$10^6$  \T\B                    	    & $\sim$18 			& $ 2 \times 10^{23}$                   & 0.12     	       &    $ 5 \times 10^{-6}$   \\
\hline
\end{tabular}
\label{tab:dustrtc}
\end{table}

%%%%%%%%%%%%%%%%%%%%%%%%%%%%%%%%%%%%%%%%%%%%%%%%%%%%%%%%%%%%%%%%%%%%

\section{Comparison between different tracer properties}
\label{dust-gas}
 
In this section, we compare the dust density distribution to the CO density to investigate if they are coherent and if they rise from the same medium. 
Our comparison focuses on position A in both regions where the radiative transfer calculations were carried out for both dust and gas. 
We further compare the derived dust and CO temperatures and finally present the derived values for the thermal pressure.

\subsection{Density}\label{density}

We obtain for NGC 7023 NW a density of $5 \times 10^4$ to $10^6$~${\rm cm^{-3}}$ and for NGC 7023 E of $10^4$ to $10^5$~${\rm cm^{-3}}$ using both dust and gas analysis. 
We obtain the CO column density from the gas analysis, and the H column density from the dust analysis.
Assuming $N_{\rm H_2}=N_{\rm H}/2$, we derive CO to H$_2$ abundance ratios varying between $3 \times 10^{-5}$ and $1.6 \times 10^{-4}$ (see Tab. \ref{tab:dustrtc}). 
This is in agreement with our first assumption about the CO to H$_2$ abundance ratio when we calculated the length along the line of sight from the RADEX results (see Sec. \ref{SED-7023NW}).
We can now conclude that for both dust and gas analysis we derive a similar length along the lines of sight for the densities of $5 \times 10^4 - 10^6~{\rm cm^{-3}}$ for NGC 7023 NW and $10^4 - 10^5~{\rm cm^{-3}}$ for NGC 7023 E.
This strongly suggests that dust and gas trace the same region.
It is unlikely to have two separate regions with different densities, since in all cases the region with the higher density provides enough gas and dust emission to reproduce the observations.

Additionally, using the dust density profiles we find that the CO emission occurs at an $A_{\rm V}^e$ of about 1$-$4, except for a density of around $10^6$~${\rm cm^{-3}}$ in NGC 7023 E where we find that the CO emission appears at an improbable large visual extinction.
For NGC 7023 NW, including the density profiles ($n_0=10^5$ and $10^6$ ${\rm cm^{-3}}$) in the Meudon PDR code \citep{le-petit-et-al-2006}, we find that the C/CO transition occurs at a visual extinction of ${A_V}$=2.5$-$3.5 where CO $5\le{\rm J_{\rm u}}\le$15 emits strongly since the CO abundance increases and the gas temperature is still high.
For NGC 7023 E, including the density profiles ($n_0=10^4$ and $10^5$ ${\rm cm^{-3}}$), we find that the C/CO transition occurs at a visual extinction of ${A_V}$=2$-$2.5. 
This adds new constraints on the gas density, which agrees with the above discussions.

For NGC 7023 NW, the analysis of the main atomic cooling lines ([OI] at 145 $\mu$m and [CII]), which are observed with {\it Herschel} PACS will be presented in \citet{bernard-salas-et-al-2014}. The dust density distribution derived here and included in PDR codes can reproduce the observed oxygen line emission.

\begin{center}
\begin{figure*}[t]
\includegraphics[width=0.99\textwidth]{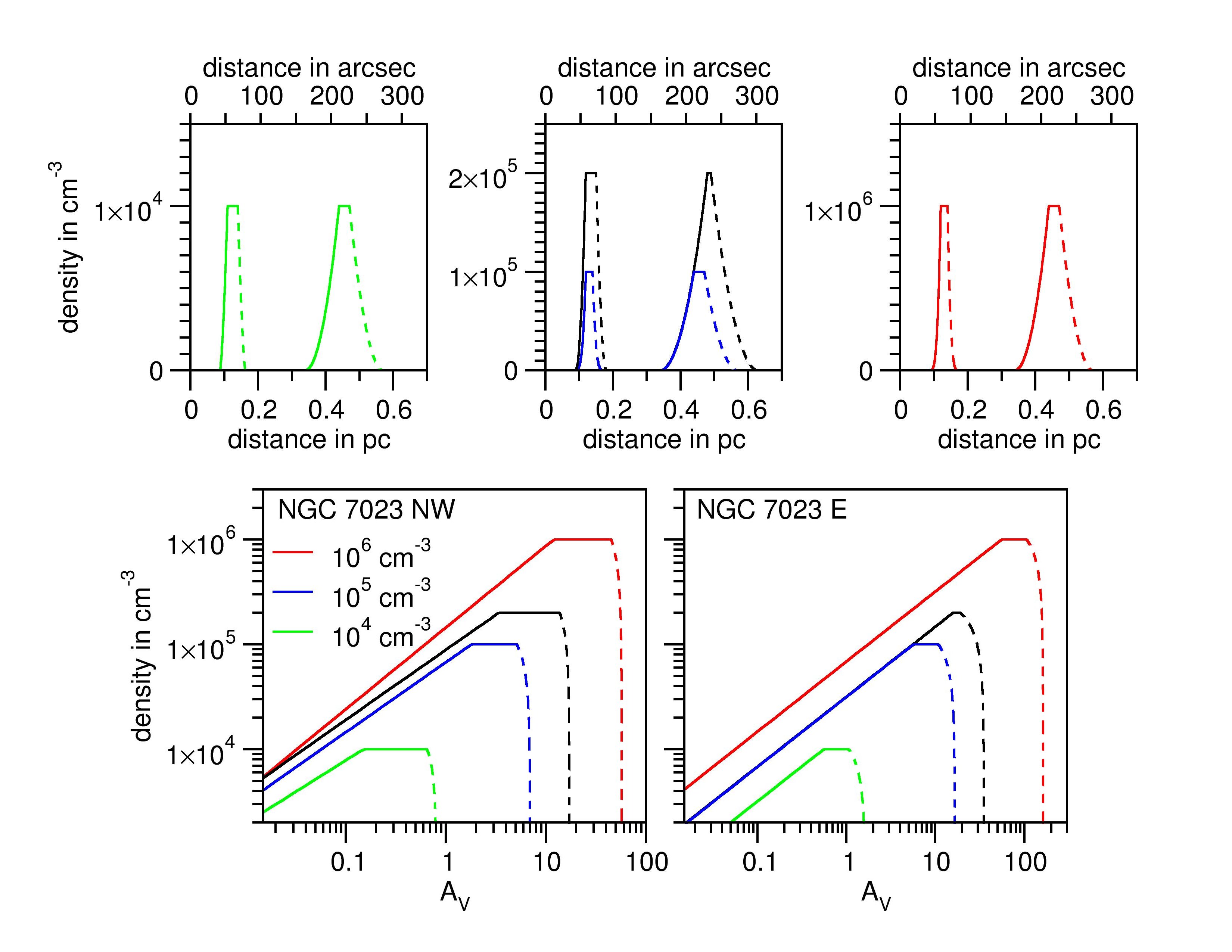}
\caption[]{The dust density profile (top) for NGC 7023 NW and NGC 7023 E (left and right curve in each plot, respectively) versus the distance from the star in pc and arcsec and versus visual extinction ${A_V}$ from the edge of the cloud (bottom) as a result from the radiative transfer calculations, left: $n_0=10^4$ ${\rm cm^{-3}}$, middle: $n_0=10^5$ ${\rm cm^{-3}}$, right: $n_0=10^6$ ${\rm cm^{-3}}$. The black curves show the density profile for the best fit shown in Fig. \ref{fig:dustres}. Since the slope of the decreasing density is ambiguous, we present it as a dashed line.}
\label{fig:denprof}
\end{figure*}
\end{center}

\begin{center}
\begin{figure*}[t]
\includegraphics[width=0.99\textwidth]{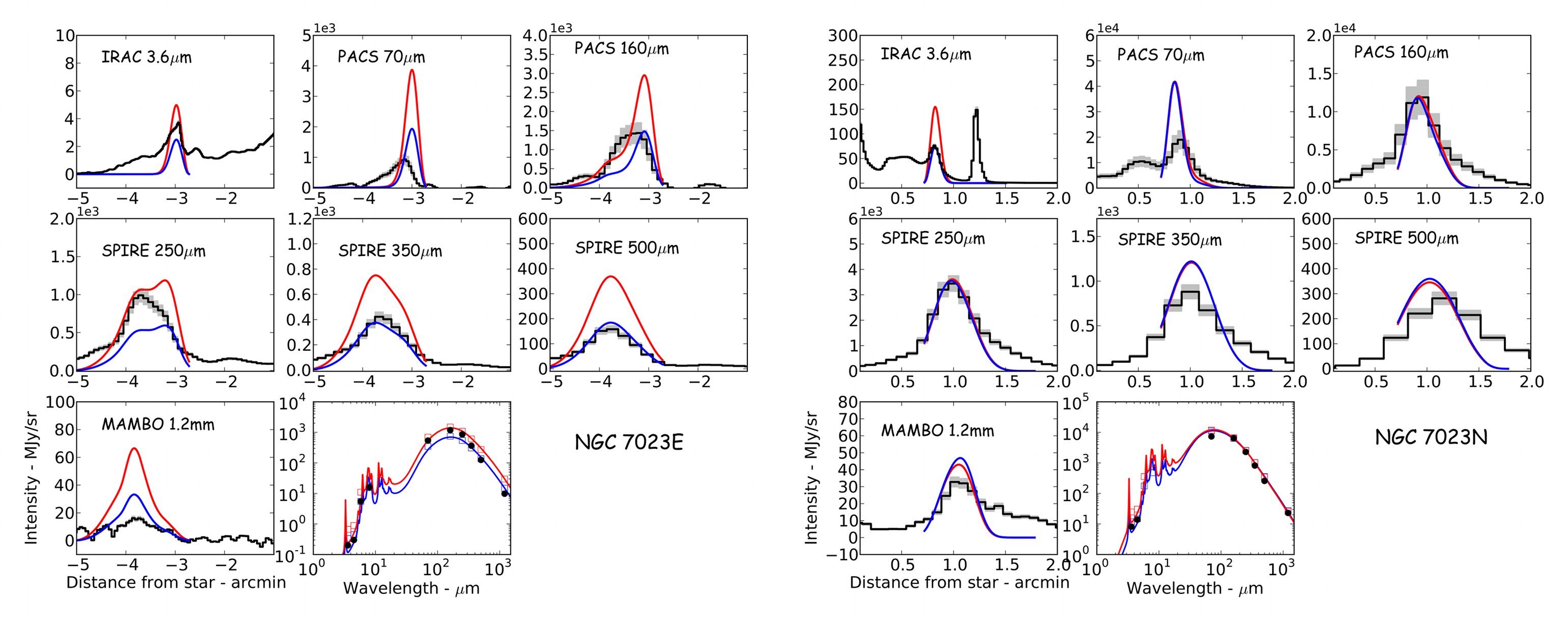}
\caption[]{The comparison of the modelled (blue and red curves) to observed data (black curves) at different wavelengths is shown. The intensity is plotted versus the angular distance from the star in arcmin. The grey zones stand for the photometric uncertainties and correspond to 5\% for IRAC, 20\% for PACS, and 15\% for SPIRE. The blue curves show the results of our model and the red curves of the model from \citet{compiegne-et-al-2011} in comparison.} 
\label{fig:dustres}
\end{figure*}
\end{center}

\begin{center}
\begin{figure}[t]
\includegraphics[width=0.45\textwidth]{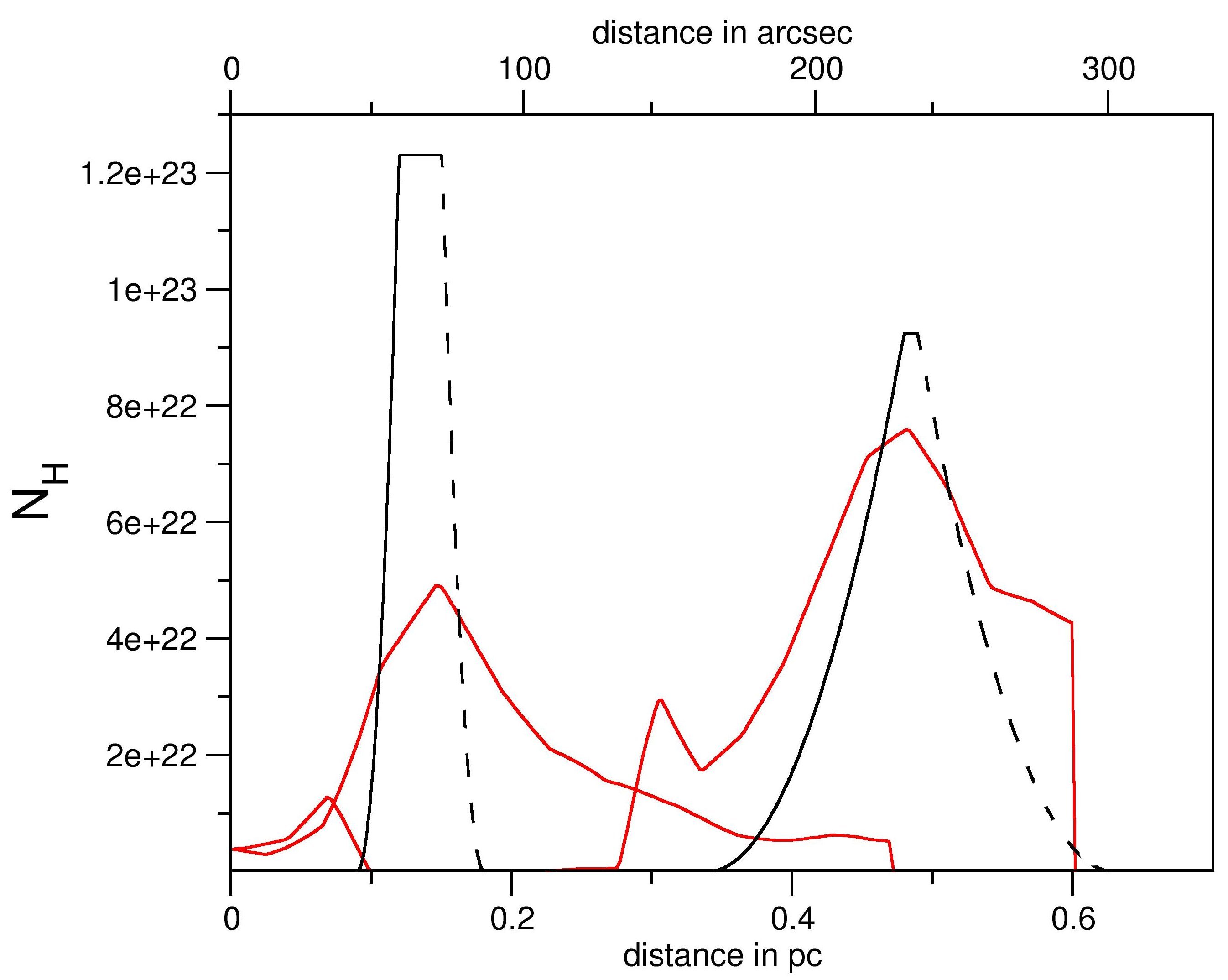}
\caption[]{The H column density profiles for NGC 7023 NW (left curves) and NGC 7023 E (right curves) derived with the black body approach (red curves) and the radiative transfer calculations (black curves). Since the slope of the decreasing density is ambiguous, we present it as a dashed line. }
\label{fig:Nh}
\end{figure}
\end{center}

\subsection{Gas and dust temperature}

We find that the gas and dust temperatures are higher close to the star and decrease with increasing distance from the star.
The gas temperatures are larger than the dust temperatures at the edge of the PDR while they are similar inside the PDR. 
For example, at positions A and C in NGC 7023 NW, we find dust temperatures of 30 K and gas temperatures of 70$-$110 K. In position B, dust temperatures are around 20 K and the gas temperatures are 30$-$50 K.
The large difference between the gas and dust temperature at the edge of the cloud is expected since in this region several processes heat the gas efficiently (e.g. photoelectric heating, H$_2$ formation) as shown in PDR models \citep{hollenbach-tielens-1999,le-bourlot-et-al-2012}.
For NGC 7023 E, the variations in gas and dust temperatures are smaller throughout the PDR and the difference between the gas and dust temperatures are smaller.
These lower temperatures and smaller variations in the dust and gas temperatures result from the smaller incident radiation field and local gas density.
Analysing different PDRs, such as the Horsehead nebula, Orion bar, NGC 2023, and $\rho$ Oph, we find that the CO excitation temperature derived from the SPIRE FTS observations, is highly dependent on the incident radiation field (Habart et al., in prep.).

\subsection{Gas thermal pressure}\label{pressure}

We derive a thermal pressure, $P=n_{\rm H} \times T$, where $n_{\rm H}$ and T are from Tab. \ref{tab:lines}, of about $6 \times10^6 - 7 \times10^7$ ${\rm K~cm^{-3}}$ at the high-J CO emission peak (position A) for NGC 7023 NW.
A high thermal pressure of $\sim 10^8$ ${\rm K~cm^{-3}}$ is also found by Joblin et al. (in prep.) by fitting the emission of $^{\rm 12}$CO J$_{\rm u}\ge15$ (PACS) data with the Meudon PDR code.
Further inside the cloud (position B) we estimate that the thermal pressure is similar while in the more diffuse region (position C), the thermal pressure is smaller by a factor of about 10.
For NGC 7023 E, we derive $P\approx1-5 \times 10^6$ ${\rm K~cm^{-3}}$ in the PDR (positions A and B), which as expected is lower than in the NW PDR.
In the cavity near NGC 7023 NW, we find a thermal pressure between $\le$$1.5 \times10^5$ and $\le$$5 \times10^5$ ${\rm K~cm^{-3}}$.
This thermal pressure is derived from the Cloudy PDR model assuming neutral gas (all ionising photons are extinguished) in the cavity with a density of $\le$$500~{\rm cm^{-3}}$ and a gas temperature of 300$-$1000 K.
If the gas in the cavity were mostly ionised, the gas temperature, and therefore thermal pressure, would be higher. 
The N$^+$ 122 $\mu$m and 205 $\mu$m lines are not detected, however, and the predicted intensity of the N$^+$ 205 $\mu$m is larger than the observed upper limit for an HII-region density of $>$$100~{\rm cm^{-3}}$. 
In the cavity near NGC 7023 E, the thermal pressure is $\le$$1.5 \times10^5$ ${\rm K~cm^{-3}}$ considering neutral gas in the cavity with a density of $\le$$500~{\rm cm^{-3}}$ and deriving a temperature of around 300 K.
In NGC 7023 NW, the thermal pressure at the high-J CO emission peak is a factor of $>$10 larger than in the cavity. 
In NGC 7023 E, the pressure in the PDR is $\ge$5 times larger than in the cavity.
NGC 7023 E could have reached thermal equilibrium pressure. 
The strong gradient in NGC 7023 NW suggests out of equilibrium processes.
Bern\'e et al. (in prep.) find evidence for gas evaporating from the molecular cloud of NGC 7023 NW analysing HIFI observations.

%%%%%%%%%%%%%%%%%%%%%%%%%%%%%%%%%%%%%%%%%%%%%%%%%%%%%%%%%%%%%%%%%%%%

\section{Star formation and physical conditions in the molecular cloud}
\label{star}

In this section, we compare the distribution of gas and dust and their physical conditions to the positions of known YSO candidates in NGC 7023 to see where stars are forming.
Using color-magnitude diagram techniques and based on {\it Spitzer} data, \citet{kirk-et-al-2009} identify about 32 candidates of YSOs within the NGC 7023 reflection nebula. 
The authors presume that the formation of these YSOs has been triggered by compression of the material around the reflection nebula shaped by the HD~200775 stars.
Several scenarios can trigger star formation: 1) gravitational instability at small scales in the dense shell forming a chain of YSO along the ionisation front \citep[e.g.][]{deharveng-et-al-2009}, 2) gravitational instability at larger scales forming local star formation regions with clusters of YSO \citep[e.g.][]{elmegreen-et-al-1977}, 3) gravitational instability in dense thin finger-like structures formed in a turbulent medium strongly irradiated by O stars \citep[e.g.][]{gritschneder-et-al-2010}, and 4) gravitational instability from pre-existent condensations compressed by the hot gas \citep[e.g.][]{bisbas-et-al-2009}.

In NGC 7023, {\citet{kirk-et-al-2009} observe that the majority of the YSO candidates are situated to the north of HD~200775 in the region coincident with the area of the high dust emission.
We exclude the scenarios 1) and 3) since no chains of YSO along the ionisation front and no finger-like structures are observed.
The scenarios 2) and 4) are more likely. 
In particular, the dust emission peak at 500 $\mu$m (position B in NGC 7023 NW) appears to coincide with the condensation seen in the IRAM-MAMBO2 image (Fig. \ref{fig:full} middle, right) and one embedded YSO (2101328+6811202, \#34 in Fig. \ref{fig:250mic}).
This YSO, not seen in the visible map, is assumed to be a probable Class 0 source, however, we cannot resolve this YSO in the SPIRE FTS maps.
The location of the YSO is far away from the ionisation front suggesting the star-formation scenario 4).
At this position we obtain a gas temperature of around 30$-$50 K and dust temperature of around 20 K, a high CO column density of around $1 \times 10^{18}~{\rm cm^{-2}}$, a ${\rm H_2}$ column density of $4.9 \times 10^{22}~{\rm cm^{-2}}$ derived from the dust and a density between $10^5$ and $10^6$ ${\rm cm^{-3}}$. Due to the low spatial resolution and the integration along the line of sight, the derived parameters are not directly associated with the YSO but rather its environment.

Two embedded YSOs (2101264+6810385 and 2101271+6810380, \#27 and \#137 in Fig. \ref{fig:250mic}, respectively) lie 20'' further inside the dust and CO emission peak (position A) in NGC 7023 NW. 
However, the YSOs, observed in the visible (Fig. \ref{fig:full}), could be between the molecular cloud and the observer and could be older than HD 200775.
They have a bolometric temperature of 1000 and 1400~K, making them probable Class II sources. 
Regarding the dust density profile, these YSOs appear to be located near the plateau in the decreasing slope if they are associated with the molecular cloud.
Analysing the observed ${\rm ^{12}CO}$ and ${\rm ^{13}CO}$ ladders with RADEX at these positions, we derive for a density between $10^5$ and $10^6$ ${\rm cm^{-3}}$, a CO column density of $2 \times 10^{18}$ ${\rm cm^{-2}}$, a ${\rm H_2}$ column density of $2 \times 10^{22}~{\rm cm^{-2}}$ derived from the dust, a gas temperature of around 30$-$50 K and a dust temperature of around 30 K. 
These derived dust and gas temperatures are not too high so that star formation is not prevented by thermal pressure.
However, we should be careful since we derive the average physical conditions along the line of sight.

It is noticeable that the NGC 7023 E region appears to be starless down to the luminosity limit of $L_{\rm bol}$=0.06~$L_{\odot}$ \citep{kirk-et-al-2009}.
In this PDR, we obtain a gas and dust temperature of around 50 K and 20 K, respectively.
The column density is high but the local gas density seems to be smaller than in NGC 7023 NW.

%%%%%%%%%%%%%%%%%%%%%%%%%%%%%%%%%%%%%%%%%%%%%%%%%%%%%%%%%%%%%%%%%%%%

\section{Conclusions}

We investigate the emission of both dust and molecular gas from the reflection nebula NGC 7023 by analysing {\it Herschel} observations with radiative transfer codes.
The NGC 7023 nebula contains three PDRs and in this study we focus on the PDRs in the E and NW directions. 
The SPIRE FTS spectroscopic cubes give us access to a wealth of intermediate ${\rm ^{12}}$CO and ${\rm ^{13}}$CO lines.
Fully sampled maps at high spectral resolution for each wavelength are analysed. 
The spectral energy distribution of CO is obtained using constant beam size data.
Using photometric data (PACS and SPIRE) combined with {\it Spitzer} data we also study the dust continuum emission.
We use radiative transfer codes for dust and gas separately to obtain the physical conditions of the emitting zones independently and to investigate if they spatially coincide.
Our main results are summarised as follows:

\begin{enumerate}

\item Spatial morphology: 
{\it Herschel} SPIRE FTS and photometer data allow us to spatially resolve the emission of dust and gas at the edge of molecular clouds.
Dust emission and CO lines show different spatial morphologies.
Dust emission is extended peaking in irradiated and/or high column density regions. 
It also reveals complex filament structures with condensations in the nebula.
CO lines show a different distribution as a function of their excitation levels. 
Low-J ${\rm ^{12}}$CO lines are rather extended, while intermediate-J and high-J ${\rm ^{12}}$CO and ${\rm ^{13}}$CO are localised probing regions of high column density of warm and dense gas.
In the PDRs, we find a spatial correlation of warm dust, ${\rm ^{12}}$CO and ${\rm ^{13}}$CO lines.

\item Density structure: 
In the PDR we are able to derive the main trend of the density profiles, column density, and spatial extent.
Since we find similar physical properties for warm dust and CO, we conclude that their emission comes from the same medium.
The density profile yields a structure with a small width of 0.05 pc and 0.2 pc and a maximum density of $5 \times 10^4$ to $10^6$ ${\rm cm^{-3}}$ and $10^4$ to $10^5$ ${\rm cm^{-3}}$ for NGC 7023 NW and NGC 7023 E, respectively.
The length along the line of sight is between the width and the extent of the emission.
The H$_2$ total column density is $\sim2 \times 10^{22}$ ${\rm cm^{-2}}$ for both PDRs.
The CO to H$_2$ abundance ratio is about $5 \times 10^{-5}-10^{-4}$ assuming that the length along the line of sight of dust and gas is equal.
The CO emission appears to occur at ${\rm A_V}$ of 1$-$4, {which corresponds to the C/CO transition as predicted by PDR codes.} 

\item Temperature and pressure:
The CO excitation temperature is 65$-$130 K (50 K) and the {fitted} dust temperature is 30 K (20 K) for NGC 7023 NW (NGC 7023 E) emission peak.
Throughout the cloud, both gas and dust temperatures decrease with increasing distance from the star. 
The dust temperature ranges from 15$-$50 K.
At the PDR edge the gas temperatures are, as expected, larger than the dust temperatures while they are similar inside the PDR.
We derive a thermal pressure in the range of $1-7 \times10^7$ ${\rm K~cm^{-3}}$ for NGC 7023 NW and $1-5 \times 10^6$ ${\rm K~cm^{-3}}$ for NGC 7023 E.
For NGC 7023 E, the thermal pressure in the cavity is slightly smaller so that this PDR might be in equilibrium, while for NGC 7023 NW the thermal pressure in the cavity is significantly lower.
Compression of the gas may have been induced by the stellar impact.
It is also possible that dense structures were present prior to the star formation.

\item Star formation: 
We compare the distribution of gas and dust and their physical conditions to the positions of known YSO candidates in NGC 7023.
In NGC 7023 NW, four candidates of YSOs are located in the region of study or along the line of sight.
The position of one YSO appears to be in agreement with the cold and dense region in the northern part of the PDR {($n_{\rm H}=10^5-10^6$ ${\rm cm^{-3}}$, $T_{\rm g}=30-50$ K, $T_{\rm d}\approx20$ K)} where the dust emission peaks at around 500 $\mu$m.
Two other YSOs are located behind the dust and CO emission peak.

\end{enumerate}

%%%%%%%%%%%%%%%%%%%%%%%%%%%%%%%%%%%%%%%%%%%%%%%%%%%%%%%%%%%%%%%%%%%%
\begin{acknowledgements} 
We thank the anonymous referee for very helpful suggestions and comments. This research acknowledges the support of the {\it Herschel} SPIRE Guaranteed Time Key project {\it Evolution of Interstellar Dust}. SPIRE has been developed by a consortium of institutes led by Cardiff Univ. (UK) and including Univ. Lethbridge (Canada); NAOC (China); CEA, LAM (France); IFSI, Univ. Padua (Italy); IAC (Spain); Stockholm Observatory (Sweden); Imperial College London, RAL, UCL-MSSL, UKATC, Univ. Sussex (UK); Caltech, JPL, NHSC, Univ. Colorado (USA). This development has been supported by national funding agencies: CSA (Canada); NAOC (China); CEA, CNES, CNRS (France); ASI (Italy); MCINN (Spain); SNSB (Sweden); STFC (UK); and NASA (USA). AF and JRG thanks the Spanish MINECO for funding support from grants CSD2009-00038 and AYA2012-32032.

\end{acknowledgements}
%%%%%%%%%%%%%%%%%%%%%%%%%%%%%%%%%%%%%%%%%%%%%%%%%%%%%%%%%%%%%%%%%%%%

\bibliographystyle{aa} % style aa.bst

\bibliography{literatur}

\begin{thebibliography}{65}
\expandafter\ifx\csname natexlab\endcsname\relax\def\natexlab#1{#1}\fi

\bibitem[{{Abergel} {et~al.}(2010){Abergel}, {Arab}, {Compi{\`e}gne}, {Kirk},
  {Ade}, {Anderson}, {Andr{\'e}}, {Baluteau}, {Bernard}, {Blagrave},
  {Bontemps}, {Boulanger}, {Cohen}, {Cox}, {Dartois}, {Davis}, {Emery},
  {Fulton}, {Gry}, {Habart}, {Huang}, {Joblin}, {Jones}, {Lagache}, {Lim},
  {Madden}, {Makiwa}, {Martin}, {Miville-Desch{\^e}nes}, {Molinari}, {Moseley},
  {Motte}, {Naylor}, {Okumura}, {Pinheiro Gon{\c c}alves}, {Polehampton},
  {Rodon}, {Russeil}, {Saraceno}, {Sauvage}, {Sidher}, {Spencer}, {Swinyard},
  {Ward-Thompson}, {White}, \& {Zavagno}}]{abergel-et-al-2010}
{Abergel}, A., {Arab}, H., {Compi{\`e}gne}, M., {et~al.} 2010, A\&A, 518, L96

\bibitem[{{Alecian} {et~al.}(2008){Alecian}, {Catala}, {Wade}, {Donati},
  {Petit}, {Landstreet}, {B{\"o}hm}, {Bouret}, {Bagnulo}, {Folsom}, {Grunhut},
  \& {Silvester}}]{alecian-et-al-2008}
{Alecian}, E., {Catala}, C., {Wade}, G.~A., {et~al.} 2008, MNRAS, 385, 391

\bibitem[{{Altamore} {et~al.}(1980){Altamore}, {Baratta}, {Cassatella},
  {Grasdalen}, {Persi}, \& {Viotti}}]{altamore-et-al-1980}
{Altamore}, A., {Baratta}, G.~B., {Cassatella}, A., {et~al.} 1980, A\&A, 90,
  290

\bibitem[{{An} \& {Sellgren}(2003)}]{an-sellgren-2003}
{An}, J.~H. \& {Sellgren}, K. 2003, ApJ, 599, 312

\bibitem[{{Andr{\'e}} {et~al.}(2011){Andr{\'e}}, {Men'shchikov}, {K{\"o}nyves},
  \& {Arzoumanian}}]{andre-et-al-2011}
{Andr{\'e}}, P., {Men'shchikov}, A., {K{\"o}nyves}, V., \& {Arzoumanian}, D.
  2011, in IAU Symposium, Vol. 270, Computational Star Formation, ed.
  J.~{Alves}, B.~G. {Elmegreen}, J.~M. {Girart}, \& V.~{Trimble}, 255--262

\bibitem[{{Arab}(2012)}]{arab-et-al-2012b}
{Arab}, H. 2012, PhD thesis, Institut d’Astrophysique Spatiale

\bibitem[{{Arab} {et~al.}(2012){Arab}, {Abergel}, {Habart}, {Bernard-Salas},
  {Ayasso}, {Dassas}, {Martin}, \& {White}}]{arab-et-al-2012}
{Arab}, H., {Abergel}, A., {Habart}, E., {et~al.} 2012, A\&A, 541, A19

\bibitem[{{Arzoumanian} {et~al.}(2013){Arzoumanian}, {Andr{\'e}}, {Peretto}, \&
  {K{\"o}nyves}}]{arzoumanian-et-al-2013}
{Arzoumanian}, D., {Andr{\'e}}, P., {Peretto}, N., \& {K{\"o}nyves}, V. 2013,
  A\&A, 553, A119

\bibitem[{{Ayasso} {et~al.}(2012){Ayasso}, {Rodet}, \&
  {Abergel}}]{ayasso-et-al-2012}
{Ayasso}, H., {Rodet}, T., \& {Abergel}, A. 2012, Inverse Problems, 28, 125005

\bibitem[{{Bedijn} \& {Tenorio-Tagle}(1981)}]{bedijn-et-al-1981}
{Bedijn}, P.~J. \& {Tenorio-Tagle}, G. 1981, A\&A, 98, 85

\bibitem[{{Benisty} {et~al.}(2013){Benisty}, {Perraut}, {Mourard}, {Stee},
  {Lima}, {Le Bouquin}, {Borges Fernandes}, {Chesneau}, {Nardetto},
  {Tallon-Bosc}, {McAlister}, {Ten Brummelaar}, {Ridgway}, {Sturmann},
  {Sturmann}, {Turner}, {Farrington}, \& {Goldfinger}}]{benisty-et-al-2013}
{Benisty}, M., {Perraut}, K., {Mourard}, D., {et~al.} 2013, A\&A, 555, A113

\bibitem[{{Bernard-Salas} {et~al.}(2014){Bernard-Salas}, {Habart}, {K\"ohler},
  {Abergel}, {Arab}, \& et~al.}]{bernard-salas-et-al-2014}
{Bernard-Salas}, J., {Habart}, E., {K\"ohler}, M., {et~al.} 2014, A\&A,
  submitted

\bibitem[{{Bern{\'e}} {et~al.}(2008){Bern{\'e}}, {Joblin}, {Rapacioli},
  {Thomas}, {Cuillandre}, \& {Deville}}]{berne-et-al-2008}
{Bern{\'e}}, O., {Joblin}, C., {Rapacioli}, M., {et~al.} 2008, A\&A, 479, L41

\bibitem[{{Bisbas} {et~al.}(2009){Bisbas}, {W{\"u}nsch}, {Whitworth}, \&
  {Hubber}}]{bisbas-et-al-2009}
{Bisbas}, T.~G., {W{\"u}nsch}, R., {Whitworth}, A.~P., \& {Hubber}, D.~A. 2009,
  A\&A, 497, 649

\bibitem[{{Compi{\`e}gne} {et~al.}(2011){Compi{\`e}gne}, {Verstraete}, {Jones},
  {Bernard}, {Boulanger}, {Flagey}, {Le Bourlot}, {Paradis}, \&
  {Ysard}}]{compiegne-et-al-2011}
{Compi{\`e}gne}, M., {Verstraete}, L., {Jones}, A., {et~al.} 2011, A\&A, 525,
  A103+

\bibitem[{{Deharveng} {et~al.}(2009){Deharveng}, {Zavagno}, {Schuller},
  {Caplan}, {Pomar{\`e}s}, \& {De Breuck}}]{deharveng-et-al-2009}
{Deharveng}, L., {Zavagno}, A., {Schuller}, F., {et~al.} 2009, A\&A, 496, 177

\bibitem[{{Elmegreen} \& {Lada}(1977)}]{elmegreen-et-al-1977}
{Elmegreen}, B.~G. \& {Lada}, C.~J. 1977, ApJ, 214, 725

\bibitem[{{Finkenzeller}(1985)}]{finkenzeller-1985}
{Finkenzeller}, U. 1985, A\&A, 151, 340

\bibitem[{{Fleming} {et~al.}(2010){Fleming}, {France}, {Lupu}, \&
  {McCandliss}}]{fleming-et-al-2010}
{Fleming}, B., {France}, K., {Lupu}, R.~E., \& {McCandliss}, S.~R. 2010, ApJ,
  725, 159

\bibitem[{{Fuente} {et~al.}(1990){Fuente}, {Martin-Pintado}, {Bachiller}, \&
  {Cernicharo}}]{fuente-et-al-1990}
{Fuente}, A., {Martin-Pintado}, J., {Bachiller}, R., \& {Cernicharo}, J. 1990,
  A\&A, 237, 471

\bibitem[{{Fuente} {et~al.}(1996){Fuente}, {Martin-Pintado}, {Neri}, {Rogers},
  \& {Moriarty-Schieven}}]{fuente-et-al-1996}
{Fuente}, A., {Martin-Pintado}, J., {Neri}, R., {Rogers}, C., \&
  {Moriarty-Schieven}, G. 1996, A\&A, 310, 286

\bibitem[{{Fuente} {et~al.}(2000){Fuente}, {Martin-Pintado},
  {Rodriguez-Fern{\'a}ndez}, {Cernicharo}, \& {Gerin}}]{fuente-et-al-2000}
{Fuente}, A., {Martin-Pintado}, J., {Rodriguez-Fern{\'a}ndez}, N.~J.,
  {Cernicharo}, J., \& {Gerin}, M. 2000, A\&A, 354, 1053

\bibitem[{{Fuente} {et~al.}(1999){Fuente}, {Mart{\'{\i}}n-Pintado},
  {Rodr{\'{\i}}guez-Fern{\'a}ndez}, {Rodr{\'{\i}}guez-Franco}, {de Vicente}, \&
  {Kunze}}]{fuente-et-al-1999}
{Fuente}, A., {Mart{\'{\i}}n-Pintado}, J., {Rodr{\'{\i}}guez-Fern{\'a}ndez},
  N.~J., {et~al.} 1999, ApJL, 518, L45

\bibitem[{{Fuente} {et~al.}(1998){Fuente}, {Martin-Pintado},
  {Rodriguez-Franco}, \& {Moriarty-Schieven}}]{fuente-et-al-1998}
{Fuente}, A., {Martin-Pintado}, J., {Rodriguez-Franco}, A., \&
  {Moriarty-Schieven}, G.~D. 1998, A\&A, 339, 575

\bibitem[{{Gerin} {et~al.}(1998){Gerin}, {Phillips}, {Keene}, {Betz}, \&
  {Boreiko}}]{gerin-et-al-1998}
{Gerin}, M., {Phillips}, T.~G., {Keene}, J., {Betz}, A.~L., \& {Boreiko}, R.~T.
  1998, ApJ, 500, 329

\bibitem[{{Gorti} \& {Hollenbach}(2002)}]{gorti-hollenbach-2002}
{Gorti}, U. \& {Hollenbach}, D. 2002, ApJ, 573, 215

\bibitem[{{Griffin} {et~al.}(2010){Griffin}, {Abergel}, {Abreu}, {Ade},
  {Andr{\'e}}, {Augueres}, {Babbedge}, {Bae}, {Baillie}, {Baluteau}, {Barlow},
  {Bendo}, {Benielli}, {Bock}, {Bonhomme}, {Brisbin}, {Brockley-Blatt},
  {Caldwell}, {Cara}, {Castro-Rodriguez}, {Cerulli}, {Chanial}, {Chen},
  {Clark}, {Clements}, {Clerc}, {Coker}, {Communal}, {Conversi}, {Cox},
  {Crumb}, {Cunningham}, {Daly}, {Davis}, {de Antoni}, {Delderfield}, {Devin},
  {di Giorgio}, {Didschuns}, {Dohlen}, {Donati}, {Dowell}, {Dowell}, {Duband},
  {Dumaye}, {Emery}, {Ferlet}, {Ferrand}, {Fontignie}, {Fox}, {Franceschini},
  {Frerking}, {Fulton}, {Garcia}, {Gastaud}, {Gear}, {Glenn}, {Goizel},
  {Griffin}, {Grundy}, {Guest}, {Guillemet}, {Hargrave}, {Harwit}, {Hastings},
  {Hatziminaoglou}, {Herman}, {Hinde}, {Hristov}, {Huang}, {Imhof}, {Isaak},
  {Israelsson}, {Ivison}, {Jennings}, {Kiernan}, {King}, {Lange}, {Latter},
  {Laurent}, {Laurent}, {Leeks}, {Lellouch}, {Levenson}, {Li}, {Li},
  {Lilienthal}, {Lim}, {Liu}, {Lu}, {Madden}, {Mainetti}, {Marliani}, {McKay},
  {Mercier}, {Molinari}, {Morris}, {Moseley}, {Mulder}, {Mur}, {Naylor},
  {Nguyen}, {O'Halloran}, {Oliver}, {Olofsson}, {Olofsson}, {Orfei}, {Page},
  {Pain}, {Panuzzo}, {Papageorgiou}, {Parks}, {Parr-Burman}, {Pearce},
  {Pearson}, {P{\'e}rez-Fournon}, {Pinsard}, {Pisano}, {Podosek}, {Pohlen},
  {Polehampton}, {Pouliquen}, {Rigopoulou}, {Rizzo}, {Roseboom}, {Roussel},
  {Rowan-Robinson}, {Rownd}, {Saraceno}, {Sauvage}, {Savage}, {Savini},
  {Sawyer}, {Scharmberg}, {Schmitt}, {Schneider}, {Schulz}, {Schwartz},
  {Shafer}, {Shupe}, {Sibthorpe}, {Sidher}, {Smith}, {Smith}, {Smith},
  {Spencer}, {Stobie}, {Sudiwala}, {Sukhatme}, {Surace}, {Stevens}, {Swinyard},
  {Trichas}, {Tourette}, {Triou}, {Tseng}, {Tucker}, {Turner}, {Vaccari},
  {Valtchanov}, {Vigroux}, {Virique}, {Voellmer}, {Walker}, {Ward}, {Waskett},
  {Weilert}, {Wesson}, {White}, {Whitehouse}, {Wilson}, {Winter}, {Woodcraft},
  {Wright}, {Xu}, {Zavagno}, {Zemcov}, {Zhang}, \&
  {Zonca}}]{griffin-et-al-2010}
{Griffin}, M.~J., {Abergel}, A., {Abreu}, A., {et~al.} 2010, A\&A, 518, L3

\bibitem[{{Gritschneder} {et~al.}(2010){Gritschneder}, {Burkert}, {Naab}, \&
  {Walch}}]{gritschneder-et-al-2010}
{Gritschneder}, M., {Burkert}, A., {Naab}, T., \& {Walch}, S. 2010, ApJ, 723,
  971

\bibitem[{{Habart} {et~al.}(2011){Habart}, {Abergel}, {Boulanger}, {Joblin},
  {Verstraete}, {Compi{\`e}gne}, {Pineau Des For{\^e}ts}, \& {Le
  Bourlot}}]{habart-et-al-2011}
{Habart}, E., {Abergel}, A., {Boulanger}, F., {et~al.} 2011, A\&A, 527, A122

\bibitem[{{Habing}(1968)}]{habing-1968}
{Habing}, H.~J. 1968, BAIN, 19, 421

\bibitem[{{Hennebelle}(2013)}]{hennebelle-et-al-2013}
{Hennebelle}, P. 2013, A\&A, 556, A153

\bibitem[{{Hollenbach} \& {Tielens}(1999)}]{hollenbach-tielens-1999}
{Hollenbach}, D.~J. \& {Tielens}, A.~G.~G.~M. 1999, Reviews of Modern Physics,
  71, 173

\bibitem[{{Hosokawa} \& {Inutsuka}(2006)}]{hosokawa-inutsuka-2006}
{Hosokawa}, T. \& {Inutsuka}, S.-i. 2006, ApJ, 646, 240

\bibitem[{{Joblin} {et~al.}(2010){Joblin}, {Pilleri}, {Montillaud}, {Fuente},
  {Gerin}, {Bern{\'e}}, {Ossenkopf}, {Le Bourlot}, {Teyssier}, {Goicoechea},
  {Le Petit}, {R{\"o}llig}, {Akyilmaz}, {Benz}, {Boulanger}, {Bruderer},
  {Dedes}, {France}, {G{\"u}sten}, {Harris}, {Klein}, {Kramer}, {Lord},
  {Martin}, {Martin-Pintado}, {Mookerjea}, {Okada}, {Phillips}, {Rizzo},
  {Simon}, {Stutzki}, {van der Tak}, {Yorke}, {Steinmetz}, {Jarchow},
  {Hartogh}, {Honingh}, {Siebertz}, {Caux}, \& {Colin}}]{joblin-et-al-2010}
{Joblin}, C., {Pilleri}, P., {Montillaud}, J., {et~al.} 2010, A\&A, 521, L25

\bibitem[{{Jones} {et~al.}(2013){Jones}, {Fanciullo}, {K{\"o}hler},
  {Verstraete}, {Guillet}, {Bocchio}, \& {Ysard}}]{jones-et-al-2013}
{Jones}, A.~P., {Fanciullo}, L., {K{\"o}hler}, M., {et~al.} 2013, \aap, 558,
  A62

\bibitem[{{Kirk} {et~al.}(2009){Kirk}, {Ward-Thompson}, {Di Francesco},
  {Bourke}, {Evans}, {Mer{\'{\i}}n}, {Allen}, {Cieza}, {Dunham}, {Harvey},
  {Huard}, {J{\o}rgensen}, {Miller}, {Noriega-Crespo}, {Peterson}, {Ray}, \&
  {Rebull}}]{kirk-et-al-2009}
{Kirk}, J.~M., {Ward-Thompson}, D., {Di Francesco}, J., {et~al.} 2009, ApJS,
  185, 198

\bibitem[{{Kreysa} {et~al.}(1998){Kreysa}, {Gemuend}, {Gromke}, {Haslam},
  {Reichertz}, {Haller}, {Beeman}, {Hansen}, {Sievers}, \&
  {Zylka}}]{kreysa-et-al-1998}
{Kreysa}, E., {Gemuend}, H.-P., {Gromke}, J., {et~al.} 1998, in Society of
  Photo-Optical Instrumentation Engineers (SPIE) Conference Series, Vol. 3357,
  Advanced Technology MMW, Radio, and Terahertz Telescopes, ed. T.~G.
  {Phillips}, 319--325

\bibitem[{{Launay} \& {Roueff}(1977)}]{launay-roueff-1977}
{Launay}, J.~M. \& {Roueff}, E. 1977, A\&A, 56, 289

\bibitem[{{Le Bourlot} {et~al.}(2012){Le Bourlot}, {Le Petit}, {Pinto},
  {Roueff}, \& {Roy}}]{le-bourlot-et-al-2012}
{Le Bourlot}, J., {Le Petit}, F., {Pinto}, C., {Roueff}, E., \& {Roy}, F. 2012,
  A\&A, 541, A76

\bibitem[{{Le Bourlot} {et~al.}(1999){Le Bourlot}, {Pineau des For{\^e}ts}, \&
  {Flower}}]{le-bourlot-et-al-1999}
{Le Bourlot}, J., {Pineau des For{\^e}ts}, G., \& {Flower}, D.~R. 1999, MNRAS,
  305, 802

\bibitem[{{Le Petit} {et~al.}(2006){Le Petit}, {Nehm{\'e}}, {Le Bourlot}, \&
  {Roueff}}]{le-petit-et-al-2006}
{Le Petit}, F., {Nehm{\'e}}, C., {Le Bourlot}, J., \& {Roueff}, E. 2006, ApJS,
  164, 506

\bibitem[{{Lemaire} {et~al.}(1996){Lemaire}, {Field}, {Gerin}, {Leach}, {Pineau
  des Forets}, {Rostas}, \& {Rouan}}]{lemaire-et-al-1996}
{Lemaire}, J.~L., {Field}, D., {Gerin}, M., {et~al.} 1996, A\&A, 308, 895

\bibitem[{{Lemaire} {et~al.}(1999){Lemaire}, {Field}, {Maillard}, {Pineau des
  For{\^e}ts}, {Falgarone}, {Pijpers}, {Gerin}, \&
  {Rostas}}]{lemaire-et-al-1999}
{Lemaire}, J.~L., {Field}, D., {Maillard}, J.~P., {et~al.} 1999, A\&A, 349, 253

\bibitem[{{Makiwa} {et~al.}(2013){Makiwa}, {Naylor}, {Ferlet}, {Salji},
  {Swinyard}, {Polehampton}, \& {van der Wiel}}]{makiwa-et-al-2013}
{Makiwa}, G., {Naylor}, D.~A., {Ferlet}, M., {et~al.} 2013, Appl. Opt., 52,
  3864

\bibitem[{{Okada} {et~al.}(2013){Okada}, {Pilleri}, {Bern{\'e}}, {Ossenkopf},
  {Fuente}, {Goicoechea}, {Joblin}, {Kramer}, {R{\"o}llig}, {Teyssier}, \& {van
  der Tak}}]{okada-et-al-2013}
{Okada}, Y., {Pilleri}, P., {Bern{\'e}}, O., {et~al.} 2013, A\&A, 553, A2

\bibitem[{{Okamoto} {et~al.}(2009){Okamoto}, {Kataza}, {Honda}, {Fujiwara},
  {Momose}, {Ohashi}, {Fujiyoshi}, {Sakon}, {Sako}, {Yamashita}, {Miyata}, \&
  {Onaka}}]{okamoto-et-al-2009}
{Okamoto}, Y.~K., {Kataza}, H., {Honda}, M., {et~al.} 2009, \apj, 706, 665

\bibitem[{{Pilleri} {et~al.}(2012){Pilleri}, {Montillaud}, {Bern{\'e}}, \&
  {Joblin}}]{pilleri-et-al-2012}
{Pilleri}, P., {Montillaud}, J., {Bern{\'e}}, O., \& {Joblin}, C. 2012, A\&A,
  542, A69

\bibitem[{{Pogodin} {et~al.}(2004){Pogodin}, {Miroshnichenko}, {Tarasov},
  {Mitskevich}, {Chountonov}, {Klochkova}, {Yushkin}, {Manset}, {Bjorkman},
  {Morrison}, \& {Wisniewski}}]{pogodin-et-al-2004}
{Pogodin}, M.~A., {Miroshnichenko}, A.~S., {Tarasov}, A.~E., {et~al.} 2004,
  A\&A, 417, 715

\bibitem[{{Racine}(1968)}]{racine-1968}
{Racine}, R. 1968, AJ, 73, 233

\bibitem[{{Rapacioli} {et~al.}(2005){Rapacioli}, {Joblin}, \&
  {Boissel}}]{rapacioli-et-al-2005}
{Rapacioli}, M., {Joblin}, C., \& {Boissel}, P. 2005, A\&A, 429, 193

\bibitem[{{Sadavoy} {et~al.}(2013){Sadavoy}, {Di Francesco}, {Johnstone},
  {Currie}, {Drabek}, {Hatchell}, {Nutter}, {Andr{\'e}}, {Arzoumanian},
  {Benedettini}, {Bernard}, {Duarte-Cabral}, {Fallscheer}, {Friesen},
  {Greaves}, {Hennemann}, {Hill}, {Jenness}, {K{\"o}nyves}, {Matthews},
  {Mottram}, {Pezzuto}, {Roy}, {Rygl}, {Schneider-Bontemps}, {Spinoglio},
  {Testi}, {Tothill}, {Ward-Thompson}, {White}, {JCMT}, \& {Herschel Gould Belt
  Survey Teams}}]{sadavoy-et-al-2013}
{Sadavoy}, S.~I., {Di Francesco}, J., {Johnstone}, D., {et~al.} 2013, \apj,
  767, 126

\bibitem[{SPIRE Observer's~Manual(2011)}]{observermanual}
SPIRE Observer's~Manual, v.~. 2011, {HERSCHEL-HSC-DOC-0798}, accessed from
  http://herschel.esac.esa.int/Documentation.shtml

\bibitem[{{Stepnik} {et~al.}(2003){Stepnik}, {Abergel}, {Bernard}, {Boulanger},
  {Cambr{\'e}sy}, {Giard}, {Jones}, {Lagache}, {Lamarre}, {Meny}, {Pajot}, {Le
  Peintre}, {Ristorcelli}, {Serra}, \& {Torre}}]{stepnik-et-al-2003}
{Stepnik}, B., {Abergel}, A., {Bernard}, J., {et~al.} 2003, A\&A, 398, 551

\bibitem[{{Swinyard} {et~al.}(2010){Swinyard}, {Ade}, {Baluteau}, {Aussel},
  {Barlow}, {Bendo}, {Benielli}, {Bock}, {Brisbin}, {Conley}, {Conversi},
  {Dowell}, {Dowell}, {Ferlet}, {Fulton}, {Glenn}, {Glauser}, {Griffin},
  {Griffin}, {Guest}, {Imhof}, {Isaak}, {Jones}, {King}, {Leeks}, {Levenson},
  {Lim}, {Lu}, {Makiwa}, {Naylor}, {Nguyen}, {Oliver}, {Panuzzo},
  {Papageorgiou}, {Pearson}, {Pohlen}, {Polehampton}, {Pouliquen},
  {Rigopoulou}, {Ronayette}, {Roussel}, {Rykala}, {Savini}, {Schulz},
  {Schwartz}, {Shupe}, {Sibthorpe}, {Sidher}, {Smith}, {Spencer}, {Trichas},
  {Triou}, {Valtchanov}, {Wesson}, {Woodcraft}, {Xu}, {Zemcov}, \&
  {Zhang}}]{swinyard-et-al-2010}
{Swinyard}, B.~M., {Ade}, P., {Baluteau}, J.-P., {et~al.} 2010, A\&A, 518, L4

\bibitem[{{Swinyard} {et~al.}(2014){Swinyard}, {Polehampton}, {Hopwood},
  {Valtchanov}, {Lu}, {Fulton}, {Benielli}, {Imhof}, {Marchili}, {Baluteau},
  {Bendo}, {Ferlet}, {Griffin}, {Lim}, {Makiwa}, {Naylor}, {Orton},
  {Papageorgiou}, {Pearson}, {Schulz}, {Sidher}, {Spencer}, {Wiel}, \&
  {Wu}}]{swinyard-et-al-2014}
{Swinyard}, B.~M., {Polehampton}, E.~T., {Hopwood}, R., {et~al.} 2014, \mnras,
  440, 3658

\bibitem[{{Teyssier} \& {Sievers}(1999)}]{teyssier-sievers-1999}
{Teyssier}, D. \& {Sievers}, A. 1999, A Fast--Mapping Method for Bolometer
  Arrary Observations. IRAM technical report.

\bibitem[{{van der Tak} {et~al.}(2007){van der Tak}, {Black}, {Sch{\"o}ier},
  {Jansen}, \& {van Dishoeck}}]{van-der-tak-et-al-2007}
{van der Tak}, F.~F.~S., {Black}, J.~H., {Sch{\"o}ier}, F.~L., {Jansen}, D.~J.,
  \& {van Dishoeck}, E.~F. 2007, A\&A, 468, 627

\bibitem[{{van Leeuwen}(2007)}]{van-leeuwen-2007}
{van Leeuwen}, F. 2007, A\&A, 474, 653

\bibitem[{{Werner} {et~al.}(2004){Werner}, {Uchida}, {Sellgren}, {Marengo},
  {Gordon}, {Morris}, {Houck}, \& {Stansberry}}]{werner-et-al-2004}
{Werner}, M.~W., {Uchida}, K.~I., {Sellgren}, K., {et~al.} 2004, ApJS, 154, 309

\bibitem[{{Wilson}(1999)}]{wilson-1999}
{Wilson}, T.~L. 1999, Reports on Progress in Physics, 62, 143

\bibitem[{{Witt} {et~al.}(2006){Witt}, {Gordon}, {Vijh}, {Sell}, {Smith}, \&
  {Xie}}]{witt-et-al-2006}
{Witt}, A.~N., {Gordon}, K.~D., {Vijh}, U.~P., {et~al.} 2006, ApJ, 636, 303

\bibitem[{{Wu} {et~al.}(2013){Wu}, {Polehampton}, {Etxaluze}, {Makiwa},
  {Naylor}, {Salji}, {Swinyard}, {Ferlet}, {van der Wiel}, {Smith}, {Fulton},
  {Griffin}, {Baluteau}, {Benielli}, {Glenn}, {Hopwood}, {Imhof}, {Lim}, {Lu},
  {Panuzzo}, {Pearson}, {Sidher}, \& {Valtchanov}}]{wu-et-al-2013}
{Wu}, R., {Polehampton}, E.~T., {Etxaluze}, M., {et~al.} 2013, \aap, 556, A116

\bibitem[{{Yang} {et~al.}(2010){Yang}, {Stancil}, {Balakrishnan}, \&
  {Forrey}}]{yang-et-al-2010}
{Yang}, B., {Stancil}, P.~C., {Balakrishnan}, N., \& {Forrey}, R.~C. 2010, ApJ,
  718, 1062

\bibitem[{{Yuan} {et~al.}(2013){Yuan}, {Wu}, {Li}, {Yu}, \&
  {Miller}}]{yuan-et-al-2013}
{Yuan}, J.-H., {Wu}, Y., {Li}, J.~Z., {Yu}, W., \& {Miller}, M. 2013, MNRAS,
  429, 954

\bibitem[{{Zylka}(1998)}]{zylka-1998}
{Zylka}, R. 1998, Pocket Cookbook for MOPSIC Software.

\end{thebibliography}

\end{document}